\documentclass[floatfix, prb, twocolumn, showpacs]{revtex4}

\usepackage{graphicx}

\usepackage{dcolumn}

\usepackage{bm}

\begin{document}

\preprint{\today}
\title{Low energy excitations and dynamic Dzyaloshinskii-Moriya 
interaction in $\alpha'$-NaV$_2$O$_5$
studied by far infrared spectroscopy
}

\author{T.~R\~o\~om}
\email{roomtom@kbfi.ee}

\author{D.~H\"uvonen}
\author{U.~Nagel}

\affiliation{National
Institute of Chemical Physics and Biophysics,
    Akadeemia tee 23, 12618 Tallinn, Estonia.}
\author{Y.-J.~Wang}
\affiliation{National High Magnetic Field Laboratory,
Florida State University,
1800 East Paul Dirac Drive,
Tallahassee, FL 32306
}

\author{R.\,K.~Kremer}
\affiliation{Max-Planck-Institut f\"ur Festk\"orperforschung,
Heisenbergstra\ss e 1, D-70569 Stuttgart, Germany}

\date{\today  \\ cc Accepted 
for publication in Physical Review B }

\begin{abstract}

We have studied far infrared transmission spectra of $\alpha'$-NaV$_2$O$_5$ 
between 3 and  200\,cm$^{-1}$ 
in polarizations of incident light parallel to $a$, $b$, and $c$ crystallographic axes
in magnetic fields up to 33\,T. 
The temperature dependence of the transmission spectra  was studied close to and below 
the phase transition temperature $T_c=34$\,K.
The triplet origin of an excitation  at 65.4\,cm$^{-1}$ (8.13\,meV) is revealed 
by splitting  in the magnetic field.
The g-factors for the triplet state are $g_a=1.96\pm0.02$,
$g_b=1.975\pm0.004$ and $g_c=1.90\pm0.03$.
The magnitude of the spin gap at low temperatures is found to be magnetic field independent 
at least up to 33\,T. 
All other infrared-active transitions appearing below $T_c$ 
are ascribed to zone-folded phonons. Two different dynamic Dzyaloshinskii-Moriya (DM) 
mechanisms 
have been discovered that contribute to the oscillator strength 
of the otherwise forbidden singlet to triplet transition.
	\textit{First}, 
the strongest singlet to triplet transition is an electric dipole transition
where the polarization of the incident light's electric field is
parallel to the ladder rungs ($\mathbf{E}_1 \! \parallel \! \mathbf{a} $).
This electric dipole active transition is allowed by the dynamic DM interaction
created by a high frequency optical a-axis phonon.
	\textit{Second}, 
in the incident light polarization perpendicular to
the  ladder planes ($\mathbf{E}_1 \! \parallel \! \mathbf{c} $)
an  enhancement of the singlet to triplet transition is observed 
when the applied magnetic field shifts the singlet to triplet resonance frequency
to match the 68\,cm$^{-1}$ c-axis phonon energy.
The origin of  the second mechanism 
is the dynamic DM interaction created by the 68\,cm$^{-1}$ c-axis optical phonon.
The strength of the dynamic DM is calculated for both mechanisms 
using the presented theory.

\end{abstract}


\maketitle

\section{Introduction}

The opening of a spin gap is of fundamental interest in one-dimensional spin one-half systems.
In one-dimensional Heisenberg
spin chains the coupling between the spins and lattice 
leads to the spin-Peierls instability; the atomic distances change
together with the nearest neighbor exchange coupling between the spins.
As a result the spin gap opens separating the singlet ground state from the excited triplet state.
The spin-Peierls instability was discovered in organic compounds and later on in 
 inorganic CuGeO$_3$\cite{Hase1993}.
Although  $\alpha'$-NaV$_2$O$_5$ is another spin one-half quasi-one-dimensional compound 
where the spin gap opening
and the lattice distortion take place simultaneously\cite{Isobe1996,Fujii1997},
it is different from canonical spin-Peierls systems. 
The magnetic field dependence of the phase transition temperature\cite{Schnelle1999,Bompadre2000} 
$T_c$ and  the entropy change\cite{Johnston2000} at $T_c$  
are not consistent with the magneto-elastically driven phase transition in $\alpha'$-NaV$_2$O$_5$, 
where in addition to the displacement of atoms 
at $T_c=34$\,K
a new charge order appears\cite{Nakao2000,Grenier2002,Ohama1999,Revurat2000}. 
An extra degree of freedom comes from one electron
being shared by the V-O-V rung
as $\alpha'$-NaV$_2$O$_5$ is a quarter-filled two leg spin ladder compound.
In the low temperature phase unpaired electrons on V-O-V rungs 
shift from the middle of the rung to off-center positions and a zigzag pattern
of V$^{4+}$ and V$^{5+}$ ions along the ladder legs
exists on all ladders\cite{Nakao2000,Grenier2002,Smaalen2002}.
There are four ladder planes in the unit cell at ambient pressure
and under high pressure a series of successive phase transitions 
to phases with more than four planes in the unit cell are observed\cite{Ohwada2001Devil}.

$\alpha'$-NaV$_2$O$_5$ is not an ideal one-dimensional spin system
as revealed by inelastic neutron scattering (INS)\cite{Yosihama1998,Grenier2001}.
In addition to the dispersion of magnetic excitations along the ladder direction
two dispersion curves with a rather small dispersion of 1.2\,meV 
along the ladder rung direction are observed. 
One curve is at 8meV and the other at 10meV  at the center of the Brillouin zone. 
More exact spin gap value, 8.13\,meV (65.5\,cm$^{-1}$), has been determined 
by the high field electron spin resonance\cite{Luther1998,Nojiri2000}.

Doubling of the lattice constants along a- and b-axes and quadrupling
along c-axis  create additional
Raman and infrared active modes at the phase transition temperature.
The question is whether they are all zone-folded lattice modes
or some of them are spin excitations.
The controversial modes are infrared active modes polarized
along the c-axis   at 68 and 106\,cm$^{-1}$
and Raman modes with frequencies nearly matching 
the frequencies of  the two infrared modes.
The origin of the 68 and 106\,cm$^{-1}$ excitations  is of fundamental interest
since  the 68\,cm$^{-1}$ excitation is almost degenerate with the 65.5\,cm$^{-1}$ triplet.
Spin chain models do not predict a bound singlet excitation 
being degenerate with the triplet excitation\cite{Uhrig1996,Bouzerar1998,Zheng2001}.

Electric and magnetic dipole transitions between singlet and
triplet states are forbidden in principle.
The reason for this is the different parity of the ground singlet state and  the excited triplet state.
The singlet state is anti-symmetric and  the  triplet state is symmetric
relative to the interchange of two spins. 
Electric dipole or magnetic dipole operators, responsible for the optical absorption,
will not couple these two states.
An anti-symmetric interaction can  mix singlet and triplet states.
If such interaction exists the   transitions are partially allowed.
The strength  of the partially allowed optical transition,
either magnetic dipole or electric dipole,
depends on the orientation of the light polarization and  applied magnetic field with 
respect to  the crystal axes.
Using infrared spectroscopy it is possible not only to extract
a separation of  energy levels in  a spin system, 
but also find the origin of the transition, either electric dipole or magnetic dipole, and 
the orientation of the anti-symmetric interaction. 

\begin{figure}[tb] 
    \includegraphics[width=8.6cm]{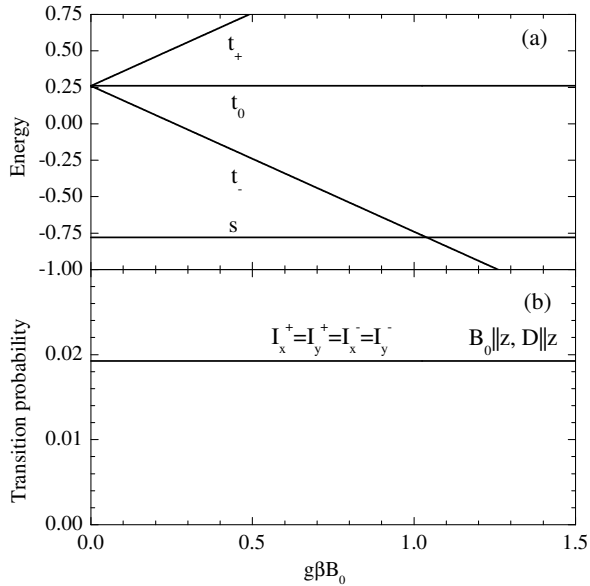}
\caption{Two spins coupled by the isotropic exchange coupling, $J=1$, and the
DM interaction, $ \mathbf{D}  \parallel  \mathbf{B}_0, D=0.4 $.
(a) -- energy levels; $ |\mathrm{t}_- \rangle = | \mathrm{T}_-\rangle $ and 
$ |\mathrm{t}_+ \rangle = | \mathrm{T}_+\rangle $ 
are  pure triplet states in any field.
(b) -- non-zero transition probabilities 
$I_i^j$ from the ground state $|\mathrm s \rangle$ to the triplet state $|\mathrm t_j \rangle$
for a given orientations $i=x\ \mathrm{or}\  y$ of the alternating magnetic field $\mathbf{H}_1$.
}
\label{MagDipHpar}
\end{figure}

In this paper we study spin gap excitations and phonons  in $\alpha'$-NaV$_2$O$_5$
using far-infrared spectroscopy.
To explain the experimentally observed singlet to triplet absorption
we present  a calculation of the dynamic Dzyaloshinskii-Moriya (DM) 
absorption mechanism
by numerical diagonalization of the spin-phonon hamiltonian.
Analytical results in the perturbation theory for the small 
dynamic DM interaction are given.
Also, the theory of the second relevant mechanism,
the magnetic dipole active static DM mechanism  is presented.

\begin{figure}[tb]
    \includegraphics[width=8.6cm]{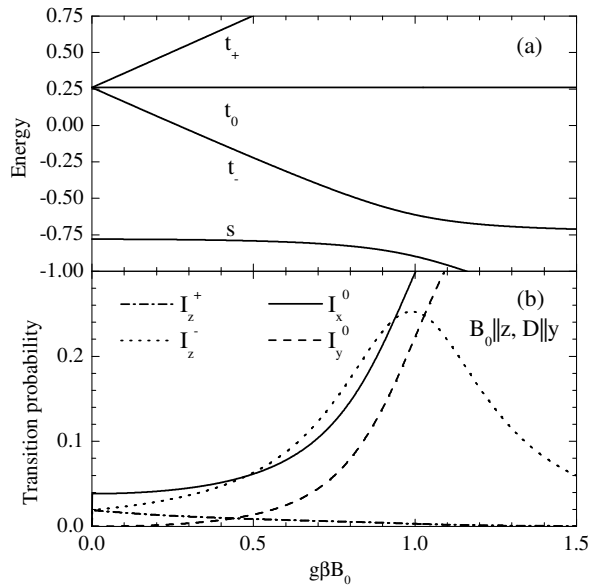}
\caption{Two spins coupled by the isotropic exchange coupling, $J=1$, and the
DM interaction, $ \mathbf{D} \perp \mathbf{B}_0, D=0.4 $.
(a) -- energy levels; $| \mathrm{t}_0\rangle = | \mathrm{T}_0\rangle $ is a pure triplet state in any field.
(b) -- non-zero transition probabilities $I_i^j$ from the ground state 
$|\mathrm s \rangle$ to the triplet state $|\mathrm t_j \rangle$
for given orientations $i=x, y, \mathrm{or}\  z$ of the alternating magnetic field $\mathbf{H}_1$.
}
\label{MagDipHperp}
\end{figure}

\section{Static and dynamic Dzyaloshinskii-Moriya interaction}

The anti-symmetric  DM interaction,
introduced by Dzyaloshinskii\cite{Dzhyaloshinskii1958} and Moriya\cite{Moriya1960}, 
is  a combination of  superexchange and spin-orbital interactions
and is linear in spin-orbital coupling. 
For a particular spin system the allowed components of the DM interaction are
determined by the symmetry of the spin complex\cite{Dzhyaloshinskii1958,Moriya1960}.
Corrections to the energy spectrum due to the DM interaction are usually small
because the correction is proportional to $D^2/\Delta$, 
where $D$ is the magnitude of the DM interaction
and $\Delta$ is the separation of the singlet and triplet energy levels.
In addition to the DM interaction there is a symmetric spin-spin interaction 
that is quadratic in spin-orbital coupling.
Although the symmetric interaction does not couple the singlet and triplet states
it affects the splitting of the triplet state sublevels.
It was shown by Shekhtman 
\textit{et al.}\cite{Shekhtman1992,Shekhtman1993} 
for a single bond superexchange that
the triplet state remains degenerate in zero magnetic field
if both symmetric and anti-symmetric spin-spin interactions are taken into 
account.

Although the corrections to the energy levels are small,
the mixing of the singlet and the triplet state by 
the anti-symmetric interaction could be enough to produce experimentally 
detectable optical singlet to triplet transitions. 
The transition probabilities for the magnetic dipole operator in spin chains with DM interactions
were calculated by Sakai \textit{ et al.} \cite{Sakai2000}.
The  idea that the electric dipole singlet to triplet transition is partially allowed
when an optically active phonon lowers the crystal symmetry and therefore creates 
a dynamic DM interaction, was put forward by C{\'e}pas \textit{ et al.}\cite{Cepas2001,Cepas2002}. 
Below we calculate the energy spectra and transition probabilities 
using a simple two spin model.
In the case of  the static DM mechanism the
full spin hamiltonian with Shekhtman corrections is used.
For the dynamic DM interaction  we extend the theory of  C{\'e}pas \textit{ et al.}
beyond the perturbation theory and solve the hamiltonian by exact diagonalization.

\subsection{Magnetic dipole transitions  and Dzyaloshinskii-Moriya interaction 
for a spin pair}\label{MagDipTheory}

In this section we calculate the energy spectrum and
the magnetic dipole transition probabilities 
for two isotropically exchange coupled spins  ($S=1/2$) in the presence 
of the anti-symmetric DM interaction and the second-order symmetric interaction.
We find the eigenvalues and eigenstates of the hamiltonian $H = H_0+H_{sx}$
and calculate the transition probabilities from the ground state induced by 
the magnetic dipole operator
\begin{equation}
H_{md}= g\mu _B \mathbf{H}_1\cdot (\mathbf{S}_1+\mathbf{S}_2),
\label{magdip}
\end{equation}
where $\mathbf{H}_1$ is the magnetic field component of the light, 
$g$ is the electron g-factor, and $\mu _B$ is the Bohr magneton.
The zero-order hamiltonian is
\begin{equation}
H_0=J \mathbf{S}_1 \cdot  \mathbf{S}_2 + g\mu _B \mathbf{B}_0 \cdot (\mathbf{S}_1 +\mathbf{S}_2),
\label{pair_exchange}
\end{equation}
where $J$ is the isotropic exchange coupling between spins $\mathbf{S}_1 $ and $\mathbf{S}_2$, and
$\mathbf{B}_0$ is the applied static magnetic field.
The first and second order corrections
are (Eq.(2.19) from Ref.\cite{Shekhtman1993})
\begin{equation}
H_{sx} = -\frac{| \mathbf{D} |^2}{4J} \mathbf{S}_1 \cdot  \mathbf{S}_2 +
\frac{1}{2J}\mathbf{S}_1 \cdot \mathbf{D}\mathbf{D} \cdot \mathbf{S}_2 +
H_{DM},
\label{pair_sx}
\end{equation}
where $\mathbf{D} $ is the DM vector
and we have separated the anti-symmetric DM interaction 
\begin{equation}
H_{DM} = \mathbf{D} \cdot  [\mathbf{S}_1 \times   \mathbf{S}_2].
\label{pair_DM}
\end{equation}

We choose singlet and triplet as the basis of eigenstates since they 
are the eigenstates of $H_0$.
These states are the singlet 
$ | \mathrm{S} \rangle=(|+- \rangle-|-+ \rangle) / \sqrt{2} $
and the three components of the triplet 
$ | \mathrm{T}_-\rangle = | - -\rangle$,
$ | \mathrm{T}_0 \rangle=(|+- \rangle + |-+ \rangle)/ \sqrt{2} $,  and   
$ | \mathrm{T}_+\rangle = | + + \rangle$.
The spin quantization axis $z$ is chosen parallel to the applied field $\mathbf{B}_0$
and for a single spin $ \langle+ | S_z | +\rangle = - \langle - | S_z | -\rangle = 1/2 $.
Let the state vector be 
$| \Psi \rangle = (\mathrm{T}_+, \mathrm{T}_0, \mathrm{T}_{-}, \mathrm{S})$.
We  diagonalize the hamiltonian $H=H_0+H_{sx}$ 
for two orientations of applied field, along the DM vector 
and perpendicular to the DM vector, denoting the eigenstates by
$(\mathrm{t}_+, \mathrm{t}_0, \mathrm{t}_{-}, \mathrm{s})$.

$\mathbf{B}_0 \parallel \mathbf{D} \parallel \mathbf{z} $. 
The hamiltonian in the matrix representation is 
\begin{widetext}
\begin{equation}
H = \left(
\begin{array}{cccc}
\frac{1}{4}J + \frac{1}{16}D^2J^{-1} + G_z & 0 & 0 & 0 \\
0 & \frac{1}{4}J - \frac{3}{16}D^2J^{-1}  & 0 & - \frac{1}{2}\, \imath D \\
0 & 0& \frac{1}{4}J + \frac{1}{16}D^2J^{-1} - G_z & 0 \\
0 & \frac{1}{2}\, \imath D  & 0 & -\frac{3}{4}J + \frac{1}{16}D^2J^{-1}
\end{array}
\right),
\label{HparPair}
\end{equation}
\end{widetext}
where $G_z=g \mu _B B_0 $ is the Zeeman term.
For arbitrarily chosen $J=1$ and $D=0.4$ the energy levels are shown in 
Fig.\,\ref{MagDipHpar}(a).
The symmetric part of $H_{sx}$ adds a correction $D^2/(16J)$ to all energy levels
except $ |\mathrm{T}_0 \rangle $ where it is $-3D^2/(16J)$.
This correction for $ |\mathrm{T}_0 \rangle $ is partially canceled by $H_{DM}$ 
that mixes
$ |\mathrm{S} \rangle $ and  $ |\mathrm{T}_0 \rangle $.
As a result the triplet state sublevels stay degenerate in zero magnetic field
as was pointed out in ref.\cite{Shekhtman1992,Shekhtman1993}.
The net effect of $H_{sx}$ in zero field is to lower the singlet state energy
by $-3D^2/(16J)$ and to raise the triplet state energy by $D^2/(16J)$.

The transition probability for the magnetic dipole operator (\ref{magdip}) 
 from the ground state  is calculated as 
$I_i^j = | \langle \mathrm{t}_j |S_{1i}+S_{2i}|\mathrm{s} \rangle |^2$,
$i=x, y, z$. 
 The alternating magnetic field $\mathbf{H}_1$ polarized along $x$ or $y$ axis (perpendicular 
to $ \mathbf{B}_0 $ and  $ \mathbf{D} $) gives non-zero intensities 
as shown in Fig.\,\ref{MagDipHpar}(b).
This is because $ | \mathrm{s} \rangle $ has the triplet component $ | \mathrm{T}_0\rangle $
mixed in and the transitions from $ | \mathrm{T}_0\rangle $ to $ | \mathrm{T}_-\rangle$ and 
$| \mathrm{T}_+\rangle $ are allowed by $\mathrm{S}_x$ and $\mathrm{S}_y$ operators.
The transition probabilities do not depend on the strength of the applied field
since the mixing of $ | S\rangle $ and  
$ | \mathrm{T}_0\rangle $ is independent of $ \mathbf{B}_0 $.

$\mathbf{B}_0 \perp  \mathbf{D} \parallel \mathbf{y} $. 
The hamiltonian is 
\begin{widetext}
\begin{equation}
H = \left(
\begin{array}{cccc}
\frac{1}{4}J - \frac{1}{16}D^2J^{-1} + G_z & 0 & -\frac{1}{8}D^2J^{-1} & \frac{\sqrt{2}}{4}D \\
0 & \frac{1}{4}J + \frac{1}{16}D^2J^{-1} & 0 & 0 \\
-\frac{1}{8}D^2J^{-1} & 0& \frac{1}{4}J - \frac{1}{16}D^2J^{-1} - G_z & \frac{\sqrt{2}}{4}D \\
\frac{\sqrt{2}}{4}D  & 0 & \frac{\sqrt{2}}{4}D  & -\frac{3}{4}J + \frac{1}{16}D^2J^{-1}
\end{array}
\right).
\label{HperPair}
\end{equation}
\end{widetext}
In this field orientation $ | \mathrm{T}_-\rangle $ and $ | \mathrm{T}_+\rangle $ 
are mixed into $ | \mathrm{S}\rangle$ by $H_{DM}$
and $ |  \mathrm{t}_0 \rangle$ remains a pure state, 
$ |  \mathrm{t}_0 \rangle =  |  \mathrm{T}_0 \rangle$.
Note that 
there is an avoided crossing at $g\mu _B B_0\approx 1$
between $|  \mathrm{s} \rangle $ and $ |  \mathrm{t}_- \rangle $
as shown in Fig.\,\ref{MagDipHperp}(a). 

The transitions from $ |  \mathrm{s} \rangle $
to $ |  \mathrm{t}_- \rangle $ and $ |  \mathrm{t}_+\rangle $ are observed 
when $\mathbf{H}_1\parallel\mathbf{B}_0$,
see $I_z^-$ and $I_z^+$ in Fig.\,\ref{MagDipHperp}(b).
In high magnetic field $I_z^-$ prevails over $I_z^+$ because the mixing of $| \mathrm{T}_-\rangle$
into the ground state increases and the mixing of $| \mathrm{T}_+\rangle$ decreases.
Finite transition probability $ I_x^0 $ to the $ | \mathrm{t}_0\rangle $ is observed in small fields 
when $\mathbf{H}_1\perp \mathbf{B}_0, \mathbf{D}$ 
whereas $ I_y^0 = 0$ ($\mathbf{H}_1\parallel \mathbf{D}$)  as $B_0$ approaches zero.
Both transition probabilities are determined by the amount 
$|\mathrm{T}_- \rangle$ and $|\mathrm{T}_+\rangle$  are mixed into the ground state
since transition operators $\mathrm{S}_x$ and $\mathrm{S}_y$ couple these two states 
to the $|\mathrm{t}_0 \rangle = |\mathrm{T}_0 \rangle$ state. 
$ I_x^0 $ and $ I_y^0 $ gain intensity as the ground state changes into $|\mathrm{T}_- \rangle$ 
with increasing field.

In summary, the following selection rules are observed for the magnetic dipole transition 
from the singlet to the triplet state in the presence of DM interaction.
\textit{ First}, if the magnetic field is parallel to the DM vector
$\mathbf{D}\parallel \mathbf{B}_0$, transitions to the 
triplet state sublevels  $|\mathrm{t}_- \rangle$ and $|\mathrm{t}_+\rangle$  are observed. 
These transitions have  field-independent intensities and do not depend on 
polarization in the plane perpendicular to the DM vector,
$\mathbf{H}_1\perp \mathbf{D}$ .
\textit{ Second},  if the magnetic field is perpendicular to the DM vector
$\mathbf{B}_0\perp \mathbf{D}$, then in small fields
($B_0\ll J/g\mu _B$) the transition to $|\mathrm{t}_0 \rangle$ has a weak field dependence
and is observed in polarization $\mathbf{H}_1\perp \mathbf{B}_0, \mathbf{D}$.
The transitions to  $|\mathrm{t}_- \rangle$ and $|\mathrm{t}_+\rangle$  are observed
in $\mathbf{H}_1\parallel \mathbf{B}_0$ polarization. 
In this polarization in magnetic fields, $ g\mu _B B_0 \geq D$,
the transition probability to  $|\mathrm{t}_- \rangle$ increases and 
to $|\mathrm{t}_+\rangle$  decreases with increasing field.
Sakai \textit{ et al.} \cite{Sakai2000} calculated magnetic dipole transition probabilities 
for interacting spin chains using a 16 spin cluster.
In their model the symmetric anisotropic superexchange was not considered.
Our single dimer model leaves the triplet levels degenerate
whereas the degeneracy is lifted in their calculation.
Whether the degeneracy will be lifted or not when the symmetric anisotropic 
superexchange in addition to the anti-symmetric DM interaction is included 
in their model needs a separate study.

\subsection{\label{dynDMtheory}Electric dipole transitions and dynamic 
Dzyaloshinskii-Moriya interaction for a spin pair}

We  show that the electric field component of the far-infrared  light $\mathbf{E}_1$
that couples to an optically active phonon can cause transitions between singlet 
and triplet states if this phonon creates a DM interaction by lattice deformation.

Electric dipole coupling between the phonon and the light in the long wavelength limit is 
\begin{equation}
V=eQE_1,
\label{eldipoleham}
\end{equation}
where $e$ is an effective charge associated with a lattice
normal coordinate $Q$. 
Here we assume that the electric field is polarized parallel to the 
electric dipole moment of the normal coordinate $Q$
and we have dropped the time dependence of $V$.

We expand the DM vector $\mathbf{D} $  into a power series of $Q$
\begin{equation}
\mathbf{D}(Q) =  \mathbf{D}(0) + 
\frac{\partial  \mathbf{D}}{\partial Q}\left| \right._{Q=0}Q + \ldots. 
\label{series}
\end{equation}
The first term is the static DM vector in the absence of lattice deformation.
We already demonstrated in the previous section that this interaction gives rise 
to magnetic dipole transitions between singlet and triplet states. 
Here for simplicity we take $\mathbf{D}(0)=0$.
We will ignore terms quadratic in $\mathbf{D}$
in $H_{sx}$ (Eq.\ref{pair_sx}) because these  symmetric interactions
will not give us any transitions 
between singlet and triplet states.
Leaving out higher order terms of $Q$  we get for the DM interaction (\ref{pair_DM})
\begin{equation}
H_{DMQ} =Q
 \mathbf{D}_Q  \cdot  [\mathbf{S}_1 \times   \mathbf{S}_2],
\label{pair_QDM}
\end{equation}
where $ \mathbf{D}_Q \equiv  \frac{\partial  \mathbf{D}}{\partial Q}\left| \right._{Q=0} $.
For the phonon system we use the secondary quantization presentation.
The lattice normal coordinate $ Q$ can be presented in terms of 
creation and annihilation operators $a^\dagger$ and $a$, $ Q=q(a^\dagger + a)$,
where $q$ is the transformation coefficient.
Since we left out $Q^2$ and higher order terms in Eq.(\ref{series}), 
the dynamic DM interaction will couple two  phonon states 
$|n \rangle$  and $|n'\rangle$, where 
$n' \! =n \! \pm 1$; $n$ is the occupation number of phonons 
in mode $Q$.

\begin{figure}[bt]
    \includegraphics[width=8.6cm]{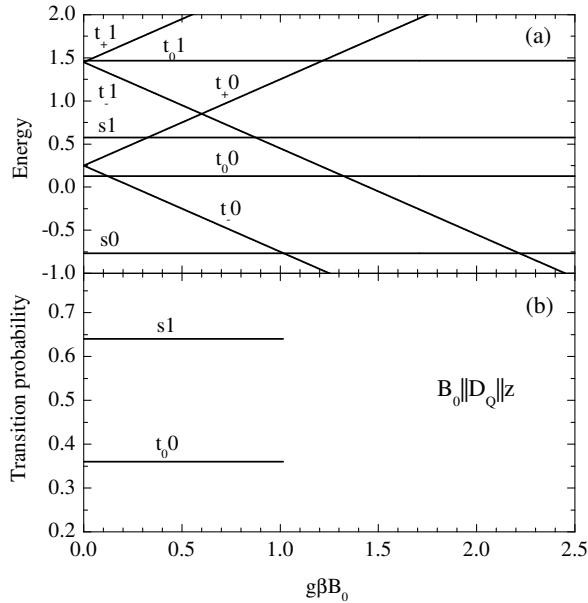}
\caption{Two spins coupled by the isotropic exchange coupling, $J=1$, and 
the dynamic DM interaction, $ qD_Q=0.4$,  created by the phonon with  
a frequency $\hbar\omega_p=1.2$;
 $\mathbf{D}_Q \parallel \mathbf{B}_0 $.
(a) -- energy levels;
$| \mathrm{t}_{\pm} 0 \rangle$ and $| \mathrm{t}_{\pm} 1 \rangle$ 
are pure states $|\mathrm{T}_{\pm}0 \rangle$ and $|\mathrm{T}_{\pm}1 \rangle$ in any field.
(b) -- non-zero electric dipole transition probabilities  from the 
ground state  to the coupled spin-phonon state
$| \mathrm{t}_0 0 \rangle$ and $| s1 \rangle$. 
The graph is not extended  above the field 
where the ground state changes from the singlet $| s0 \rangle$ 
to the triplet $|\mathrm{t}_-0 \rangle$. 
}
\label{ElDipHpar}
\end{figure}

\begin{figure}[tb]
    \includegraphics[width=8.6cm]{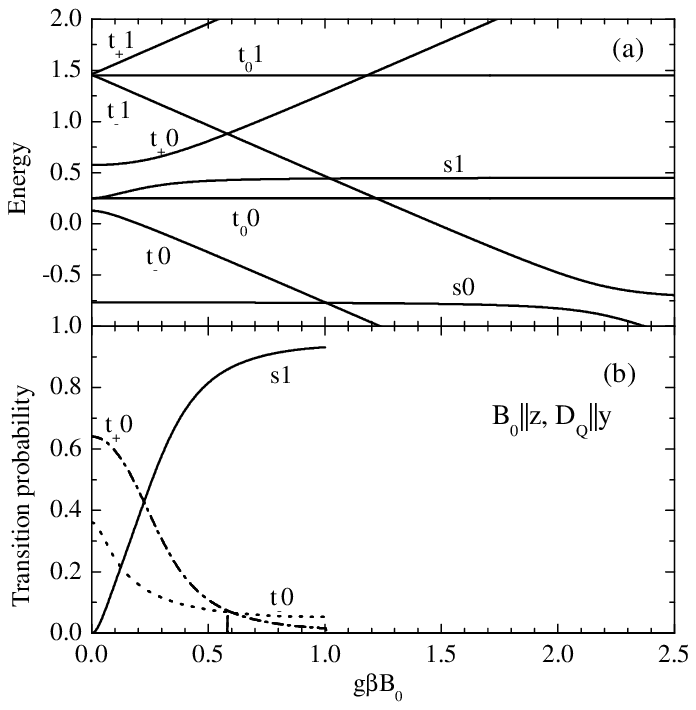}
\caption{Two spins coupled by the isotropic exchange coupling, $J=1$, and 
the dynamic DM interaction, $ qD_Q=0.4$,  created by the phonon with  
a frequency $\hbar\omega_p=1.2$;
 $\mathbf{D}_Q \perp \mathbf{B}_0 $.
(a) -- energy levels;
$| \mathrm{t}_0 0 \rangle$ and $| \mathrm{t}_01 \rangle$ are  
pure states $|\mathrm{T}_00 \rangle$ and $|\mathrm{T}_01 \rangle$ in any field.
(b) -- non-zero electric dipole transition probabilities  
from the ground state  to the coupled spin-phonon state
$| \mathrm{t}_- 0 \rangle$ (dotted), $| \mathrm{t}_+ 0 \rangle$ (dash-dotted), 
and $| s1 \rangle$ (solid).
The graph is not extended  above the field 
where the ground state changes from the singlet $| s0 \rangle$ 
to the triplet $|\mathrm{t}_-0 \rangle$. 
}
\label{ElDipHperp}
\end{figure}

The hamiltonian of the coupled spin-phonon system is
\begin{equation}
H  =  \hbar \, \omega_p \, a^\dagger a +
J \mathbf{S}_1 \cdot  \mathbf{S}_2 + 
g\mu _B \mathbf{B}_0 \cdot (\mathbf{S}_1+\mathbf{S}_2)+ H^{(1)}_{DMQ},
\label{dynamic0order}
\end{equation} \vspace{-18pt}
\begin{equation}
H^{(1)}_{DMQ} = q(a^\dagger + a) \mathbf{D}_Q  \cdot  [\mathbf{S}_1 \times   \mathbf{S}_2], 
\label{dynamic1order}
\end{equation}
$ \hbar \, \omega_p$ is the phonon energy. 
In the low temperature limit  $k_BT \! \ll \! \hbar  \omega $
the thermal population of phonon states is low, $\langle n \rangle \! \approx \! 0$, and we
can consider only the phonon states with either 0 or 1 phonon, 
$ | 0\rangle  $ and $ | 1\rangle  $.
After diagonalization of the hamiltonian (\ref{dynamic0order})
we treat $V$, Eq.(\ref{eldipoleham}), as a time-dependent 
perturbation to calculate the transition probabilities from the ground state.
We choose $ | \mathrm{S} \rangle  $  and $ | \mathrm{T}_i\rangle  $
with the  quantization axis along the applied field $\mathbf{B}_0$ as the basis  for the spin states. 
Let the state vector be
$\Psi= (\mathrm{T}_+1,  \mathrm{T}_+0, \mathrm{T}_01, \mathrm{T}_00,  
\mathrm{T}_-1,  \mathrm{T}_-0, \mathrm{S}1, \mathrm{S}0)$.
The hamiltonian is diagonal in this basis except for the last term, $H^{(1)}_{DMQ}$. 
The eigenstates are labeled by $ | \mathrm{s} n\rangle  $  and $ | \mathrm{t}_i n\rangle  $
where $n=0$ or 1.
We solve two separate cases,  $\mathbf{D}_Q\parallel \mathbf{B}_0$
and $\mathbf{D}_Q\perp \mathbf{B}_0$.

$ \mathbf{D}_Q\parallel \mathbf{B}_0\parallel z$.
In this field orientation  $ \mathbf{D}_Q = (0,0,D_Q) $.
The diagonal elements are the same as in (\ref{HparPair}) 
except that the phonon energy $\hbar\omega_p$ will be added if $n=1$.
Beside the diagonal elements the non-zero elements
of the hamiltonian (\ref{dynamic0order}) are the ones created by $ H^{(1)}_{DMQ} $:
$  \langle \mathrm{S}1 | H^{(1)}_{DMQ} |\mathrm{T}_00 \rangle = 
\langle \mathrm{S}0 | H^{(1)}_{DMQ} |\mathrm{T}_01 \rangle =
-\langle \mathrm{T}_01 | H^{(1)}_{DMQ} |\mathrm{S}0 \rangle =
 - \langle \mathrm{T}_00 | H^{(1)}_{DMQ} |\mathrm{S}1 \rangle = \imath qD_Q/2$.
The energy levels, calculated for $J=1$, $\hbar\omega_p = 1.2$, and  $ qD_Q=0.4 $,
 are shown in Fig.\,\ref{ElDipHpar}(a).
The largest repulsion is between $ |\mathrm{s}1 \rangle $ and $ |\mathrm{t}_00 \rangle  $, which
are the linear combinations of  $ |\mathrm{S}1 \rangle $ and $ |\mathrm{T}_00 \rangle  $.
The other two mixed together states
are $ |\mathrm{S}0 \rangle $ and $ |\mathrm{T}_01 \rangle $ 
giving us the ground state $ |\mathrm{s}0 \rangle $ and $ |\mathrm{t}_01 \rangle $. 
One has to keep in mind that not only the spin states are mixed, but also the phonon states
$ |0 \rangle $ and $ |1 \rangle $ are mixed.
All other 4 states that involve triplet states $ | \mathrm{T}_{\pm}\rangle $  are pure states.

The splitting of energy levels has to be taken  with some precaution. 
The splitting due to the dynamic DM is observed when there is one phonon
excited, $n=1$. 
This is not the case at thermal equilibrium at low T when $\langle n \rangle =0$
(the possible role of zero-point vibrations is ignored in our approach).
If the phonon is brought to the state $n=1$ by the  light-phonon interaction (\ref{eldipoleham})
the effect of one phonon on the shift of energy levels should be observed in the experiment,
unless it  is  much smaller than the lifetime or inhomogeneous broadening of energy levels. 
The magnitude of the shift and whether it could be  observed  in the experiment or not 
will not affect conclusions about the transition probabilities.

Calculation of the transition probability $ |\langle \mathrm{t}_j n'|V| \mathrm{s}0 \rangle|^2 $
is straightforward since $V$ couples  states that are diagonal in the basis of pure spin states
and non-diagonal in the basis of phonon states $ |0 \rangle $ and $ |1 \rangle $.
Two transitions from the ground state $ | \mathrm{s}0 \rangle $ 
are observed, to $ | \mathrm{s}1 \rangle $ and $ |\mathrm{t}_00 \rangle  $,
shown in Fig.\,\ref{ElDipHpar}(b).
If the dynamic DM is zero, then $ |\mathrm{s}0 \rangle= |\mathrm{S}0 \rangle $,  
$ |\mathrm{s}1 \rangle= |\mathrm{S}1 \rangle $,
and $ |\mathrm{t}_00 \rangle= |\mathrm{T}_00 \rangle $.
The transition from $ |\mathrm{s}0 \rangle$  to $ |\mathrm{s}1 \rangle $ is
an ordinary absorption of an infrared photon  $\hbar\omega_p =E_{\mathrm{s}1} - E_{\mathrm{s}0}$   
with probability 1, and
the transition to $|\mathrm{t}_00 \rangle $ has zero probability.
When the dynamic DM interaction is turned on, additional absorption sets in and
a photon  of energy $E_{\mathrm{t}_00} - E_{\mathrm{s}0}$ is   absorbed.
This can be viewed as a virtual excitation of a phonon by the light to the state $ | 1\rangle$
while the spin state remains singlet, and then the dynamic DM interaction (\ref{pair_QDM})
brings the (virtual) phonon back to $ | 0\rangle$ while changing the spin state to $|\mathrm{T}_0 \rangle $.
The polarization of the absorbed photon with respect to the crystal axes
is determined by the phonon states involved.

$ \mathbf{D}_Q\perp \mathbf{B}_0\parallel z$.
We take $ \mathbf{D}_Q = (0, D_Q,0) $.
Beside diagonal elements there are 8 non-zero  elements
$\langle \mathrm{T}_+1 | H^{(1)}_{DMQ} |\mathrm{S}0 \rangle$,
$\langle \mathrm{T}_+0 | H^{(1)}_{DMQ} |\mathrm{S}1 \rangle$,
$\langle \mathrm{T}_-1 |H^{(1)}_{DMQ}|\mathrm{S}0 \rangle$, 
$\langle \mathrm{T}_-0 |H^{(1)}_{DMQ}|\mathrm{S}1 \rangle$, 
$\langle \mathrm{S}1 |H^{(1)}_{DMQ}|\mathrm{T}_+0 \rangle$, 
$\langle \mathrm{S}1 |H^{(1)}_{DMQ}|\mathrm{T}_-0 \rangle$,
$\langle \mathrm{S}0 |H^{(1)}_{DMQ}|\mathrm{T}_+1 \rangle$,
$\langle \mathrm{S}0 |H^{(1)}_{DMQ}|\mathrm{T}_-1 \rangle$, all equal to
$\sqrt{2}qD_Q/4$. 
The energy levels, calculated for $J=1$, $\hbar\omega_p = 1.2$, and  $ qD_Q=0.4 $,
are plotted in Fig.\,\ref{ElDipHperp}(a).

The strongest mixing occurs between 
$ | \mathrm{S}1 \rangle$ and $ | \mathrm{T}_{\pm} 0 \rangle $
giving the eigenstates 
$ | \mathrm{s}1 \rangle$ and $ | \mathrm{t}_{\pm} 0 \rangle $.
Also, there is an additional mixing between $ |\mathrm{T}_-0 \rangle $ and $ |\mathrm{T}_+0 \rangle $ 
levels in small fields.

In small fields, $g \mu _B B_0 < J$,
the mixing of $ |\mathrm{S}0 \rangle $ and $ |\mathrm{T}_{\pm}1 \rangle $ 
is less pronounced since their separation 
is larger than the separation of $ |\mathrm{S}1 \rangle $ and $ |\mathrm{T}_{\pm}0 \rangle $. 
Therefore, for the analysis of the transition probabilities
the ground state can be taken as pure $|\mathrm{S}0 \rangle $  and 
the transition probabilities are mainly determined by the mixing
between  $|\mathrm{S}1 \rangle $ and $ |\mathrm{T}_{\pm}0 \rangle$.  
The effect of mixing of $|\mathrm{T}_{\pm}1\rangle$ into the ground state has a secondary effect on the 
transition probabilities.
Transitions from the ground state to three excited states 
$|\mathrm{t}_+0 \rangle $, $|\mathrm{t}_-0 \rangle $, and $|\mathrm{s}1 \rangle $ 
 have non-zero probabilities (see Fig.\,\ref{ElDipHperp}(b)).
The transition probability to the $|\mathrm{s}1 \rangle $ state increases with field 
because $|\mathrm{s}1 \rangle $ changes gradually from the mixed state into 
$ |\mathrm{S}1 \rangle $.
The transition probabilities to $|\mathrm{t}_+0 \rangle $ and $|\mathrm{t}_-0 \rangle $  
state decrease as the  field increases
because the amount of $| \mathrm{S} 1\rangle $ mixed into them decreases.
Again, the polarization of the absorbed photon with respect to the crystal axes
is determined by the phonon states involved.

\textit{Perturbation theory.}
Analytical results can be obtained in the limit $|E_{\mathrm{S}1}-E_{\mathrm{T}_i0}| \gg qD_Q$,
$i=-, 0, +$.
This case holds when $|\hbar \omega_p - (J \pm g\mu _B B_0)| \gg qD_Q$.
We   find the first order perturbation corrections to the states $| \mathrm{S}n \rangle$
and $| \mathrm{T}_i n \rangle$, where $n=0,1$, using $H^{(1)}_{DMQ}$, Eq.\,(\ref{dynamic1order}),
as perturbation.
Then the transition probabilities are calculated between the new states 
$ | \mathrm{s}0 \rangle $ and $| \mathrm{t}_i0 \rangle$ 
as was done in the exact treatment.

For $ \mathbf{D}_Q\parallel \mathbf{B}_0\parallel \mathbf{z}$ we get
\begin{equation}
| \langle \mathrm{t}_00 | V | \mathrm{s}0 \rangle  |^2 = I_p \frac {(qD_Q)^2 (\hbar \omega_p )^2}
{[(\hbar\omega_p)^2 - J^2]^2}\, ,
\label{par_perturbation}
\end{equation}
where $I_p=(eqE_1)^2$ is the light absorption intensity by the infrared-active phonon.
The transition probability from  $| \mathrm{s}_0 \rangle$ to the triplet level $| \mathrm{t}_0 \rangle$ is independent
of the magnetic field. 

For the perpendicular case, $ \mathbf{D}_Q\perp \mathbf{B}_0\parallel \mathbf{z}$
\begin{eqnarray}
| \langle \mathrm{t}_-0 | V | \mathrm{s}0 \rangle  |^2 = I_p \frac {(qD_Q)^2 (\hbar \omega_p )^2}
{2[(\hbar\omega_p)^2 - (J-g\mu _B B_0)^2]^2}\, ,\label{perp_perturbation_minus}\\
| \langle \mathrm{t}_+0 | V | \mathrm{s}0 \rangle  |^2 = I_p \frac {(qD_Q)^2 (\hbar \omega_p )^2}
{2[(\hbar\omega_p)^2 - (J+g\mu _B B_0)^2]^2}\, .
\label{perp_perturbation_plus}
\end{eqnarray}
If $\hbar\omega_p \ll J$  then the intensity of the transition from $| \mathrm{s}0 \rangle$ 
to $| \mathrm{t}_-0 \rangle$ increases with the magnetic field 
and decreases for the transition to $| \mathrm{t}_+0 \rangle$.
If $\hbar\omega_p \gg J$  then the intensity of the $| \mathrm{s}0 \rangle$  to 
$| \mathrm{t}_+0 \rangle$ transition increases
and of $| \mathrm{s}0 \rangle$ to $| \mathrm{t}_-0 \rangle$ decreases.
In the perturbation limit the zero field intensities of the transitions from 
$| \mathrm{s}0 \rangle$ to $| \mathrm{t}_-0 \rangle$ and $| \mathrm{t}_+0 \rangle$ are equal.

\textit{In summary}, the following selection rules are obtained
for the electric dipole transition from the 
singlet to the triplet state in the presence of the dynamic DM interaction.
{\bf{1}}. The polarization of the transition: $\mathbf{E}_1$ is parallel to the dipole moment 
of the optically active phonon that creates the dynamic DM interaction.
{\bf{2}}. The orientation of the dynamic DM vector $\mathbf{D}_Q$ is determined by the symmetry of the
lattice distortion created by the optically active phonon. 
{\bf{3}}. If $\mathbf{B}_0 \parallel \mathbf{D}_Q $ a magnetic field independent 
transition probability  to the 
triplet state sublevel $ |\mathrm{t}_0 0 \rangle $ is observed. 
{\bf{4}}. If $\mathbf{B}_0 \perp \mathbf{D}_Q $ magnetic field dependent transition probabilities  to the 
triplet state sublevels $ |\mathrm{t}_+0 \rangle $ and $ |\mathrm{t}_-0 \rangle $ are observed.

\section{Experimental}

We studied several single crystals of $\alpha^\prime$-Na$_x$V$_2$O$_y$
from the batch E106\cite{Kremer1999Loa}.
According to the heat capacity measurements these crystals have
$T_c=33.9$\,K, and
the chemical composition  $x=1.02$ and $y=5.06$.
The (ab)-plane properties,
$\mathbf{E}_1\, \bot \, \mathbf{c}$,
$\mathbf{k}\!\parallel \!\mathbf{c}$,
were studied on three single crystals,
600\,$\mu$m (area in the (ab)-plane 21\,mm$^2$), 
120\,$\mu$m (20\,mm$^2$) 
and
40\,$\mu$m (3.5\,mm$^2$) 
thick in
$\mathbf{c}$-direction.
The (bc)-plane properties,
$\mathbf{E}_1\, \bot \, \mathbf{a}$, $\mathbf{k}\!\parallel \!\mathbf{a}$,
were measured on a mosaic of three crystals,
each approximately 650\,$\mu$m thick, with a total area of 19\,mm$^2$ in the (bc)-plane.
The (ac)-plane properties,
$\mathbf{E}_1\, \bot \, \mathbf{b}$, $\mathbf{k}\!\parallel \!\mathbf{b}$,
were measured on a mosaic of 7 crystals,
each approximately 800\,$\mu$m thick, with a total area of 8.5\,mm$^2$ in the (ac)-plane.

The far-infrared measurements were done with a polarizing
Martin-Puplett Fourier transform spectrometer\cite{Sciencetech}.
Light pipes were used to guide the far infrared light into 
the sample cryostat equipped with a 12\,T Oxford Instruments magnet
and two silicon bolometers from Infrared Laboratories operated at 0.3\,K.
A rotatable polarizer was mounted at the end of the light pipe 
in front of the sample to control the polarization of light.
Spectra were recorded at  0.2 to 0.3\,cm$^{-1}$ resolution.
The magnetic  field was applied parallel to the direction of light
propagation (Faraday configuration, $\mathbf{k}\!\parallel \!\mathbf{B}_0$)
or perpendicular to the light propagation (Voigt configuration, $\mathbf{k}\,\bot \,\mathbf{B}_0$).
Measurements above 12\,T were performed at the National High Magnetic Field Laboratory
on a 33\,T Bitter magnet in the Faraday configuration using a Bruker IFS\,113v infrared spectrometer 
and a  4\,K silicon bolometer from Infrared Laboratories.

The anisotropic power absorption coefficient $\alpha_i(\omega)$ ($i=a,b,c$) was calculated
 from the measured transmission $T_i(\omega)$ 
assuming one back reflection  from the crystal front face and one from the back face,
$T_i(\omega)\!=\!(1-R_i)^2\exp[-\alpha_i(\omega)d]$,
where $d$ is the thickness of the crystal.
We used a frequency independent value for the reflectance coefficient
$R_i=[(n_i-1)/(n_i+1)]^2$.
Indexes of refraction, $n_i$, at 4K and teraherz frequencies
are $n_a=3.64$, $n_b=3.16$ and  $n_c=2.70$ (Ref.\cite{Smirnov1999}). 
According to another paper\cite{Poirier1999}  indexes of refraction  at 4K and 0.55\,cm$^{-1}$ are 
$n_a=3.07$, $n_b=3.19$ and  $n_c=2.03$.
We calculated $n_a/n_b=1.16$ from the fringe pattern in the 10 to 50\,cm$^{-1}$ range
using our transmission data.
This ratio is more similar  to the ratio $n_a/n_b=1.15$ from Ref.\cite{Smirnov1999} 
 and therefore we used their  values for refraction indexes.
The real part of the conductivity in units of $\Omega^{-1}$cm$^{-1}$ is
\begin{equation}
\sigma_1^i(\omega)=n_i(\omega)\alpha_i(\omega)/(120\pi),
\label{eq:alpha}
\end{equation}
where  in the limit of weak absorption we take $n_i$ to be independent of frequency.

\squeezetable
\begin{table}[tb]
\caption{\label{LinesTable}Absorption line positions $\omega_0$ (cm$^{-1}$), full widths at half maximum 
$\gamma $(cm$^{-1}$), and areas (cm$^{-2}$)  in 4.4\,K and 40\,K spectra of
$\alpha^\prime$-NaV$_{2}$O$_{5}$ in zero magnetic field.
Index $^T$ refers to the singlet to triplet excitation and $^F$  to the Fano line shape.}

\begin{tabular}{c|ddd|ddd}
&       & 4.4\,\textnormal{K}    &       &       & 40\,\textnormal{K}    &   \\  \cline{2-7}
& & & & & &   \\ 
& \omega_{0}  & \gamma &  \textnormal{Area}  & \omega_{0}  
& \gamma     &  \textnormal{Area}    \\
\hline & & & & & &   \\ 
$\mathbf{E}_1\!\parallel \!\mathbf{a}$
    & 65.4^{T}      & 0.6       & 6            &       &   &   \\
    & 91.2      & 0.2       & 50       &   90.7^F        & 0.8   & 120  \\
    & 101.4     & 0.26      & 110      &       &   &   \\
    & 101.7     & 0.19  & 47       &       &   &   \\
    & 111.7         & 0.18  & 8            &       &   &   \\
    & 126.7     & 0.17      & 240      &       &   &   \\
    & 127.5         & 0.22      & 140      &       &   &   \\
    & 138.5     & 0.4       & 450      & 137.7 & 2.2   & 1000  \\
    & 140          & 4.4       & 2000 & 141.0^F  &8  &5300   \\
    & 145.6     & 0.5       & 230      &       &   &   \\
    & 147.8     & 0.8       & 650      &       &   &   \\
    & 157.1     & 0.29      & 25       &       &   &   \\
    & 168.2     & 0.35      & 16       &       &   &   \\
\hline & & & & & &   \\ 
$\mathbf{E}_1\!\parallel \!\mathbf{b}$
    & 25.3      & 0.55      & 54       &       &   &   \\
    & 26.4      & 0.47      & 37       &       &   &   \\
    & 30.8      & 0.84      & 130      &       &   &   \\
    & 32.5      & 1.05      & 145      &       &   &   \\
    & 34.5      &  0.95 & 109      &       &   &   \\
    & 36.4      & 1.51      & 181      &       &   &   \\
   & 65.4^{T} & 0.2            &0.3      &       &   &   \\
    & 39.1      & 0.43      & 9            &       &   &   \\
    & 91.3      & 0.3       & 5            &       &   &   \\
    & 101.4 & 0.24      & 70       &       &   &   \\
    & 111.7 & 0.2       & 24       &       &   &   \\
    & 126.7 & 0.26      & 100      &       &   &   \\
    & 127.5 & 0.2       & 320      &       &   &   \\
    & 145.7 & 0.58      & 11       &       &   &   \\
    & 148.0 & 0.65      & 25       &       &   &   \\
    & 168.3 & 0.42  & 27       & 168.6 & 0.6   & 20   \\
    & 180      &               &               & 180  &   &   \\
    & 199.4 & 1.4       & 157      &       &   &   \\
    & 215.1 & 2.2       & 420      & 214.9 & 2.5   & 450  \\
\hline & & & & & &   \\ 
$\mathbf{E}_1\!\parallel \!\mathbf{c}$
   &65.4^{T}      & 0.2            & 0.5      &       &   &   \\
    & 68       & <1            & >100      &       &   &   \\
    & 106      & <2.5       & >110      &       &   &   \\
    & 124      &   <1          &   >50     &       &   &   \\
    & 126      &   <1.5     &   >30     &       &   &   \\
    & 130      &   <1          &   >20     &       &   &   \\
    & 132      &   <2          &   >50     &       &   &   \\

\end{tabular}
\end{table}

\section{Results}

\begin{figure}[tb]
    \includegraphics[width=8.6cm]{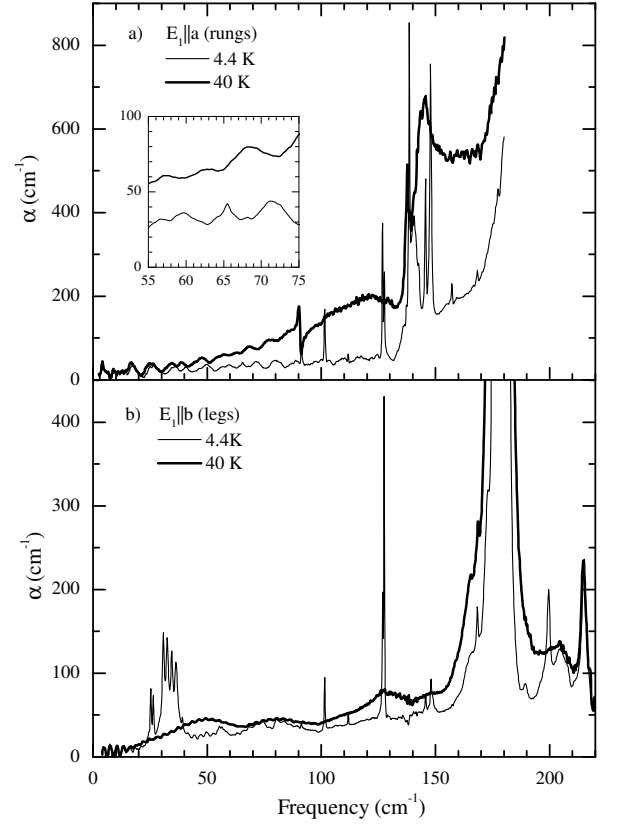}
\caption{Absorption spectra of $\alpha'$-NaV$_2$O$_5$ below, $T=4.4$\,K, and
above, $T=40$\,K, the phase transition temperature. 
(a) --   $\mathbf{E}_1\!\parallel \!\mathbf{a}$,
(b) --   $\mathbf{E}_1\!\parallel \!\mathbf{b}$.
The inset shows the singlet to triplet excitation at 65\,cm$^{-1}$.
}
\label{Fig_E_ab_4_40}
\end{figure}

\begin{figure}[tb]
    \includegraphics[width=8.6cm]{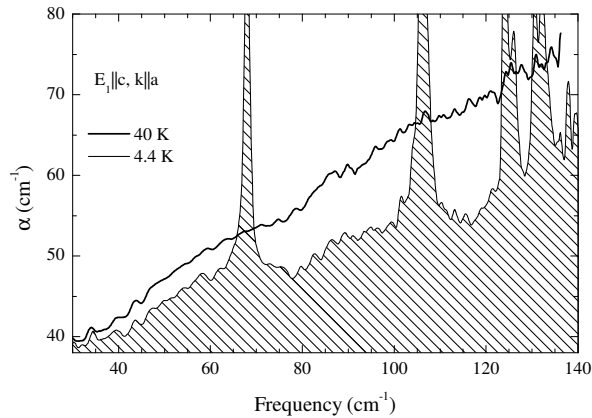}
\caption{Absorption spectra at 4.4\,K and 40\,K 
in $\mathbf{E}_1\!\parallel \!\mathbf{c}$ polarization.
}
\label{figEparC4and40}
\end{figure}

\begin{figure}[tb]
    \includegraphics[width=8.6cm]{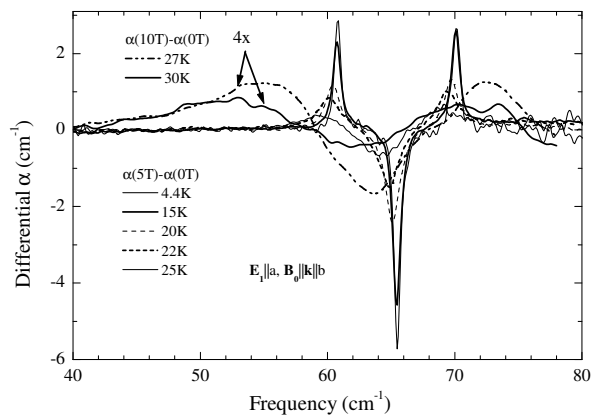}
\caption{Temperature dependence of the singlet to triplet absorption spectrum
in $\mathbf{E}_1\!\parallel  \!\mathbf{a}$ polarization.
}
\label{Spectra65Tdep2}
\end{figure}

\subsection{Absorption spectra and their temperature dependence}

The absorption spectra at temperatures  above and below
the phase transition temperature $T_c$
for  incident light polarizations $\mathbf{E}_1\!\parallel \!\mathbf{a}$ and
$\mathbf{E}_1\!\parallel \!\mathbf{b}$ are shown
in  Fig.~\ref{Fig_E_ab_4_40} and for $\mathbf{E}_1\!\parallel \!\mathbf{c}$
in Fig.~\ref{figEparC4and40}.
Below $T_c$ several new lines appear.
The line parameters are listed in Table~\ref{LinesTable}.
The full width at half maximum (FWHM) of the narrowest lines is determined by the
used instrument resolution, 0.2\,cm$^{-1}$. 
The best fit for the resolution limited narrow lines was obtained using gaussian line shapes, 
otherwise  a lorentzian line shape was used.

\begin{figure}[tb]
\includegraphics[width=8.6cm]{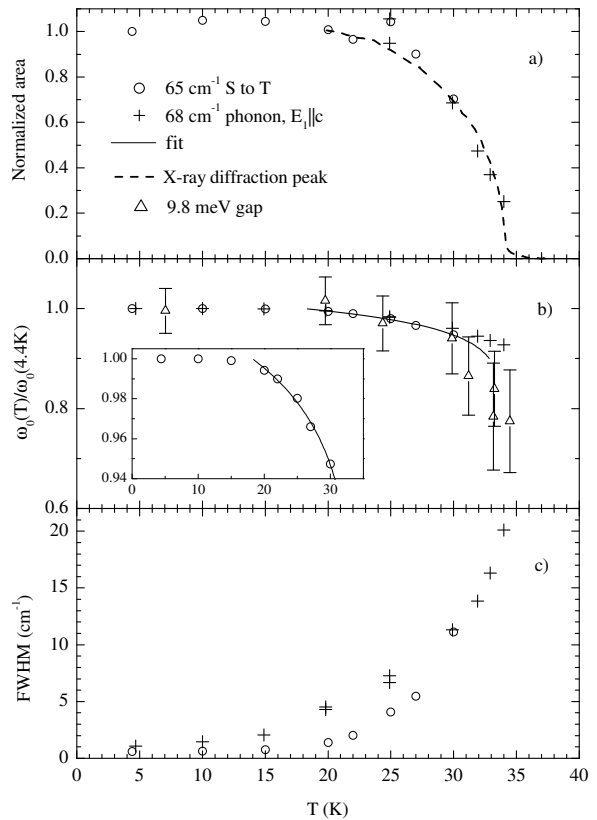} 
\caption{Zero magnetic field temperature dependence of the normalized  
absorption line area (panel a),
normalized resonance frequency (b), and FWHM (c)
for the singlet to triplet transition at 65.4\,cm$^{-1}$
$\mathbf{E}_1\!\parallel  \!\mathbf{a}$ (circles),
and for the 68\,cm$^{-1}$ c-axis phonon (crosses).
Inset to (b): the solid line is a fit of the S to T transition energy 
$\Delta (T)/\Delta (4.7{\mathrm{K}})=(1-T/T_c)^\beta$ above 20\,K; 
$T_c=33.9$\,K, $\beta=0.039\pm 0.002$.
Additionally the X-ray diffraction peak intensity 
from Ref.\cite {Gaulin2000} is plotted with a dashed line in panel (a) and
the normalized gap at 9.8\,meV measured by INS\cite{Fujii1997} is shown 
with triangles in (b).
}
\label{Tdep65_68}
\end{figure}

The $\mathbf{E}_1\!\parallel \!\mathbf{a}$ absorption at 40\,K  is
dominated by a continuous absorption  steadily increasing
towards high frequencies. Above 180\,cm$^{-1}$ the absorption
is too strong and our data is not reliable above this frequency.
There  are two derivative-like
absorption lines, one at 91.2\,cm$^{-1}$ and the other at 140\,cm$^{-1}$,
the latter being relatively broad and has its phase
opposite to the 91.2\,cm$^{-1}$ line phase.
There is a narrow line at 137.6\,cm$^{-1}$
on top of the 140\,cm$^{-1}$ line.
When $T$ is lowered, the absorption continuum  is diminished
over the entire frequency range.
Still, substantial  absorption continuum remains above    130\,cm$^{-1}$.
At $T=4$\,K the 91.2 and 140\,cm$^{-1}$ lines have
a normal absorption-like line shape. New lines appear as well:
doublets at 102, 127, 147\,cm$^{-1}$
and single lines at 112, 157, 168\,cm$^{-1}$.
The inset in Fig.~\ref{Fig_E_ab_4_40}(a) shows a weak line at
65.4\,cm$^{-1}$, the singlet to triplet excitation.
The magnetic properties of this transition are described
in more detail in section \ref{65cm}.
The $T$ dependent spectra of the singlet to triplet transition are plotted in Fig.\,\ref{Spectra65Tdep2}.
To better extract this relatively weak line the difference of 
two spectra at fixed $T$, one measured in  0\,T field and the other measured in  5
or 10\,T field, was calculated. 
One can see that the negative line, representing the zero field spectrum, together
with the positive features (the lines in 5 or 10\,T spectra) shift to lower frequencies as the temperature  increases
and at the same time the lines lose intensity and broaden.
The temperature dependence of line parameters is plotted in Fig.\,\ref{Tdep65_68}.

\begin{figure}[tb]
    \includegraphics[width=8.6cm]{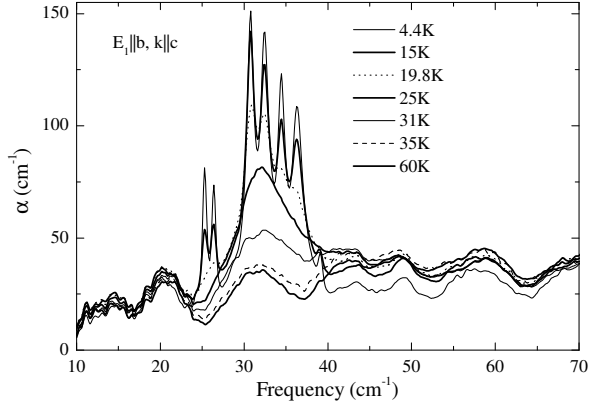}
\caption{Temperature dependence of seven low frequency  b-axis  phonons.}
\label{Multiplet2}
\end{figure}

\begin{figure}[tb]
   \includegraphics[width=8.6cm]{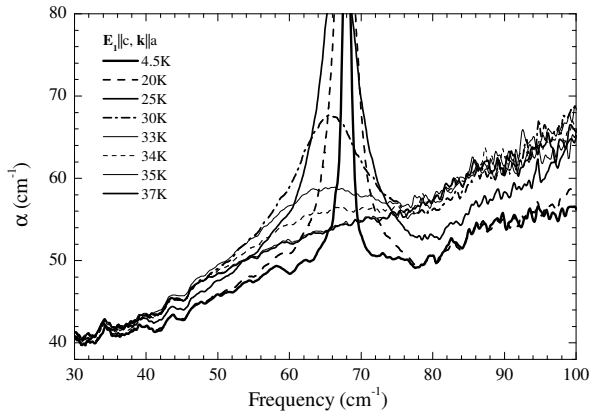}
\caption{The temperature dependence of the 68\,cm$^{-1}$ c-axis phonon absorption spectrum.}
\label{Spectra68_Oct_fig}
\end{figure}

\begin{figure}[tb]
    \includegraphics[width=8.6cm]{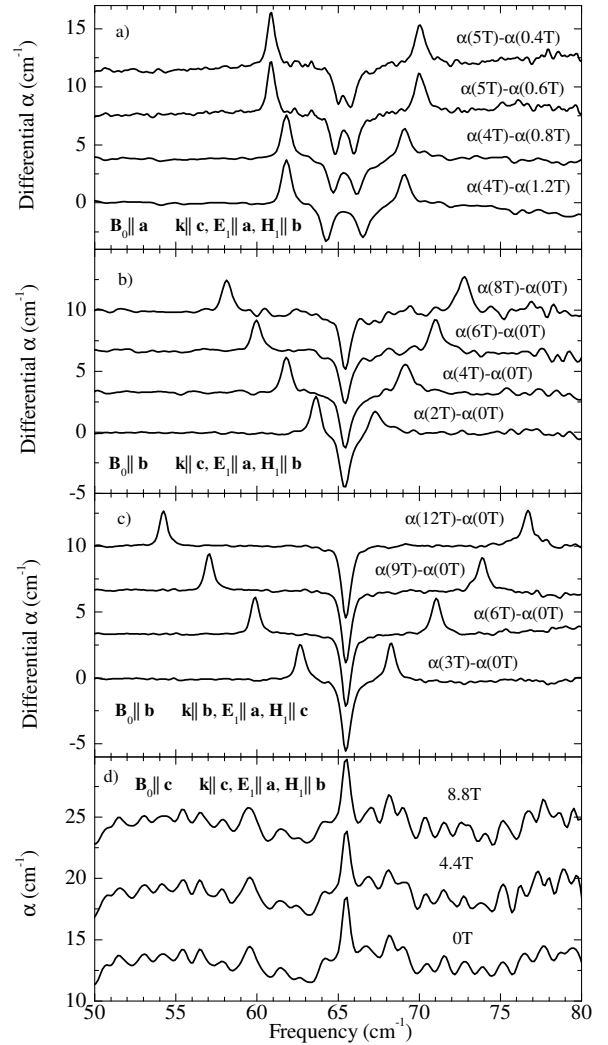}
\caption{
Magnetic field dependence of the singlet to triplet transition spectra in 
$\mathbf{E}_1 \! \parallel \! \mathbf{a}$ polarization.
Spectra have been shifted in vertical direction.
Measurements were done in Voigt (a, b) and in Faraday configuration (c, d).
}
\label{EllAfieldSpectra4}
\end{figure}

The $\mathbf{E}_1\!\parallel \!\mathbf{b}$ absorption at 40\,K is dominated
by a strong phonon line at 180\,cm$^{-1}$ and a weaker line at
215\,cm$^{-1}$, Fig.~\ref{Fig_E_ab_4_40}(b).
The temperature-dependent  absorption continuum, as was observed in
 $\mathbf{E}_1\!\parallel \!\mathbf{a}$ polarization, is absent 
in $\mathbf{E}_1\!\parallel \!\mathbf{b}$
(note the different vertical scales in Fig.~\ref{Fig_E_ab_4_40}a and
\ref{Fig_E_ab_4_40}b). 
Several lines, at
101.4, 111.7, 126.7, and 127.5\,cm$^{-1}$, have the same frequency as
in the a-axis spectrum, although different strength.
The line parameters are listed in  Table~\ref{LinesTable}.
The temperature evolution of the multiplet of seven lines 
in the $\mathbf{E}_1\!\parallel \!\mathbf{b}$ spectrum
below 40\,cm$^{-1}$  is shown in more detail in Fig.\,\ref{Multiplet2}.
Below $T_c=34$\,K a broad line appears at 32\,cm$^{-1}$.
Another line appears at 26\,cm$^{-1}$ as $T$ is lowered.
At low $T$ those two features split into a doublet and a quintet.
The line at the highest frequency, 39\,cm$^{-1}$, is relatively weak compared to other 
six lines.
The spectra shown in Fig.\,\ref{Multiplet2} have not been corrected for light interference fringes in the sample.
The 60\,K spectrum could be used as a background, but with some precaution as
some intensity is lost between 40 and 70\,cm$^{-1}$ when $T$ is lowered.

The $\mathbf{E}_1\!\parallel \!\mathbf{c}$ absorption spectrum 
has no sharp features below 140\,cm$^{-1}$ in the high temperature
phase at $T=40$\,K (Fig.~\ref{figEparC4and40}).
When $T$ is lowered below 34\,K new lines evolve at 68, 106\,cm$^{-1}$,
and two doublets around 124 and 132\,cm$^{-1}$.
Since the low $T$ transmission was too small at the transmission minimums
and  the absorption coefficient cannot be determined accurately, 
only the upper limits for the line widths and the  lower limits for 
the line areas  are given in the Table\,\ref{LinesTable}.
The temperature dependence of the 68\,cm$^{-1}$ line is shown in Fig.\,\ref{Spectra68_Oct_fig}.
The line shifts to lower frequency and broadens as $T$ is raised from 4.5\,K.
Some intensity change is still observed above $T_c$ between 34 and  35\,K, but
there are no visible differences between the 35\,K and 37\,K spectra.
The temperature dependence of line parameters is plotted in Fig.\,\ref{Tdep65_68}
together with the singlet to triplet resonance data.
Since the largest experimentally detectable absorption
 was limited by the thickness  of available  crystals
the line area has been reliably determined only above 25\,K. 
Below 25\,K the plotted FWHM  is the upper limit for the  line width.

\subsection{Singlet to triplet absorption in magnetic field}\label{65cm}

We studied the magnetic field effect on the absorption spectra
below 130\,cm$^{-1}$ at 4.4\,K
in magnetic fields up to 12\,T in all three polarizations, 
$\mathbf{E}_1\!\parallel \!\mathbf{a}$, $\mathbf{E}_1\!\parallel \!\mathbf{b}$,
and $\mathbf{E}_1\!\parallel \!\mathbf{c}$.
We did not see any line shifts nor intensity changes
except for the 65.4\,cm$^{-1}$ absorption line.

\begin{figure}[tb]
  \includegraphics[width=8.6cm]{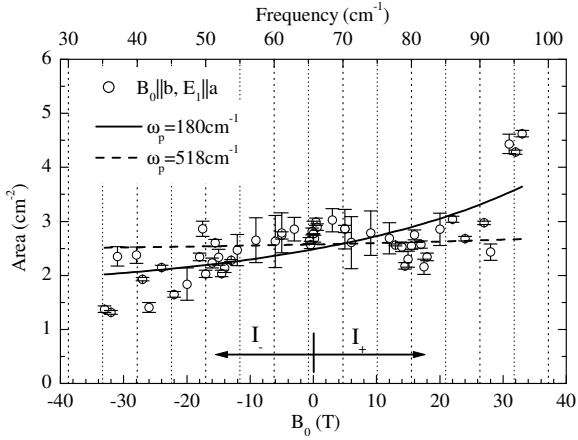}
\caption{The magnetic field, $\mathbf{B}_0\!\parallel  \!\mathbf{b}$, dependence of the singlet to triplet  
resonance line area at 4.4\,K in $\mathbf{E}_1\!\parallel  \!\mathbf{a}$ polarization.
The values for  the transition to $m_S = -1$  are plotted in the negative field direction
and for the transition to $m_S = 1$ in the positive field direction; 
$ g_a\mu _B=0.922$\,cm$^{-1}$/T.
The lines show the theoretical transition probability for the electric dipole transition
for a set of parameters where  $\omega_p$ is the resonance frequency of an optical phonon 
coupled to the spin system by the dynamic DM interaction:
solid line - $\omega_p=180$\,cm$^{-1}$,
dashed line - $\omega_p=518$\,cm$^{-1}$.
}
\label{EparA_Bdep33T}
\end{figure}

\begin{figure}[t]
    \includegraphics[width=8.6cm]{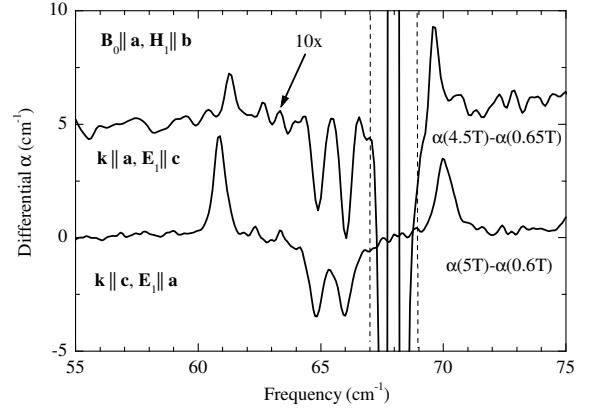}
\caption{Differential absorption spectra of the singlet to triplet 
transition at 4.4\,K in  
$\mathbf{E}_1\!\parallel  \!\mathbf{a}$ and 
$\mathbf{E}_1\!\parallel  \!\mathbf{c}$  polarizations. 
The $\mathbf{E}_1\!\parallel  \!\mathbf{c}$ spectrum has been multiplied
by 10 and offset in the vertical direction.
Both spectra have been measured in $\mathbf{B}_0\!\parallel  \!\mathbf{a}$, 
$\mathbf{H}_1\!\parallel  \!\mathbf{b}$ geometry.
}
\label{ElecDipole}
\end{figure}

For $\mathbf{E}_1\!\parallel \!\mathbf{a}$ polarization
the magnetic field effect on the 65.4\,cm$^{-1}$ line 
is demonstrated in Fig.\,\ref{EllAfieldSpectra4}.
When the field $\mathbf{B}_0$ is parallel to the a- or b-axis, the line splits (panels (a), (b), and (c)). 
In $\mathbf{B}_0\!\parallel \!\mathbf{c}$ configuration the line does not 
split or change its intensity (panel (d)).
From the field dependence of the resonance frequency the 65.4\,cm$^{-1}$
absorption line can be identified as a transition from a singlet ground state, $\mathrm{S}=0$,
to a triplet excited state, $\mathrm{S}=1$ 
Light is absorbed, depending on its polarization,
either by transitions to $m_S = 1$ and $m_S = -1$ triplet levels
when $\mathbf{B}_0\!\perp \!\mathbf{c}$
or to the magnetic field independent 
$m_S = 0$ level when $\mathbf{B}_0\!\parallel \!\mathbf{c}$.
The measurements were extended to 33\,T in one polarization and field orientation,
$\mathbf{E}_1\!\parallel \!\mathbf{a}$, $\mathbf{B}_0\!\parallel \!\mathbf{b}$.
The field dependence of the singlet to triplet transition line areas
is shown in  Fig.\,\ref{EparA_Bdep33T}. 
The lines were fitted with a lorentzian function, 
full width at half maximum (FWHM) being between 0.5 and 0.6\,cm$^{-1}$.
The line areas of the transitions to the $m_S=-1$ and $m_S=1$ levels depend weakly on the magnetic field.
Below (see section \ref{E1par_a_discussion})  we calculate the electric dipole 
transition probability using the presented theory of the dynamic DM effect, 
and compare it to our measurement results.

An important question to answer is which component of light,  $\mathbf{E}_1$ or $\mathbf{H}_1$, 
interacts with  the spin system.
What is common for the data presented in Fig.\,\ref{EllAfieldSpectra4}
is that all these measurements were done with the light polarized along the ladder rungs,
$\mathbf{E}_1\!\parallel \!\mathbf{a}$.
We made complementary measurements rotating the incident light's polarization by $90^\circ$,
thus interchanging the orientations of $\mathbf{E}_1$ and $\mathbf{H}_1$, and found that
the singlet to triplet transition was at least ten times weaker if $\mathbf{E}_1\!\perp\!\mathbf{a}$.
While the data presented in Fig.\,\ref{EllAfieldSpectra4} panels (a), (b), and (d)
still leaves open the possibility that the singlet to triplet transition could be 
a magnetic dipole transition where $\mathbf{H}_1\!\parallel \!\mathbf{b}$,
there are two other observations that contradict this.
\textit{First}, in  Fig.\,\ref{EllAfieldSpectra4} (c) $\mathbf{H}_1$ is not parallel to the b-axis
while $\mathbf{E}_1$ is still parallel to the a-axis and the transition is still strong.
\textit{Second}, in Fig.\,\ref{ElecDipole} two spectra with $\mathbf{H}_1\!\parallel \!\mathbf{b}$ 
are presented for the $\mathbf{B}_0\!\parallel \!\mathbf{a}$ applied magnetic field orientation.
The upper curve with $\mathbf{E}_1\!\parallel \!\mathbf{c}$ shows about 10 times
weaker absorption on the singlet to triplet transition than the lower curve with 
$\mathbf{E}_1\!\parallel \!\mathbf{a}$.
Since the same orientation of $\mathbf{H}_1$ gives different intensities
 the orientation of magnetic field component of light is irrelevant.
Summarizing the results presented in  Fig.\,\ref{EllAfieldSpectra4} and
\ref{ElecDipole} we conclude that the strongest contribution to the singlet to triplet absorption comes
from an electric dipole transition with the dipole moment along the a-axis.

There  are  other  contributions to the singlet to triplet absorption, 
although much weaker than in $\mathbf{E}_1\!\parallel \!\mathbf{a}$ polarization,
as is shown in Fig.\,\ref{ElecDipole}. 
The upper spectrum is for the c-axis  polarized light. 
 Two vertical dashed lines mark the area where the strong absorption due to the 
68\,cm$^{-1}$ excitation takes place in this polarization (see also Fig.\,\ref{figEparC4and40}).
One can see that the absorption lines closer to the 68\,cm$^{-1}$ excitation are 
stronger than the lines further away. 

\begin{figure}[bt]
    \includegraphics[width=8.6cm]{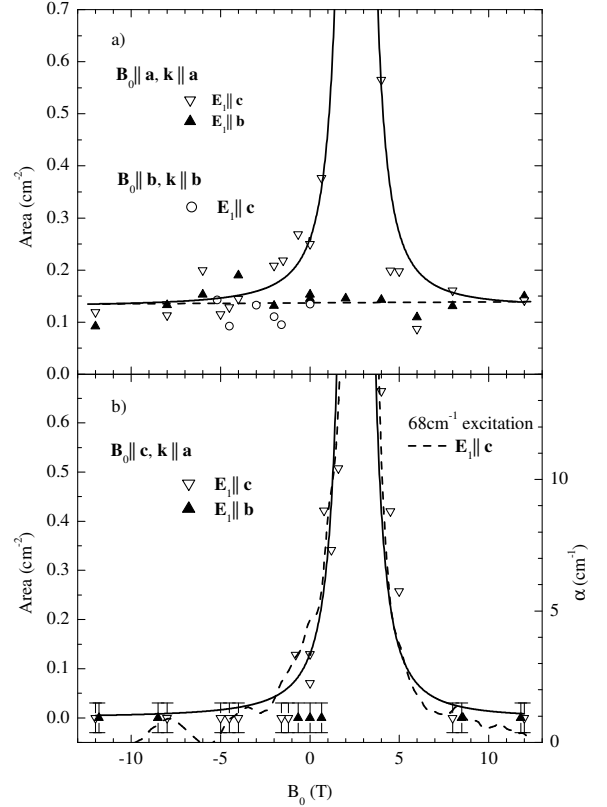}
\caption{The magnetic field dependence of the singlet to triplet transition line area at 4.4\,K
in two polarizations, $\mathbf{E}_1\!\parallel  \!\mathbf{c}$ 
(empty symbols) 
and  $\mathbf{E}_1\!\parallel  \!\mathbf{b}$ (filled triangles).  
The line areas are plotted  in the negative field direction for the
transition to the $m_S=-1$ triplet level and for the transition to $m_S=1$
in the positive field direction.
The zero field data points on the graph are one half of the measured area.
(a): $\mathbf{B}_0\!\parallel  \!\mathbf{a}$ 
and $\mathbf{B}_0\!\parallel  \!\mathbf{b}$.
(b):  $\mathbf{B}_0\!\parallel  \!\mathbf{c}$. 
The dashed line in  (a)  is a guide for the eye.
The dashed  line in  (b), units on the right axis,  
is the 68\,cm$^{-1}$ absorption line shape with the background subtracted and 
the energy units converted into magnetic field units using the triplet state g-factor, $g_c=1.90$.
The error bars shown only in (b) apply to data points in both panels.
A solid line in (b) is a fit to the dynamic  DM mechanism induced by the 68\,cm$^{-1}$ phonon.
The oscillator strength of the phonon and the singlet to triplet transition in zero field are  
400\,cm$^{-2}$ and 0.13\,cm$^{-2}$, respectively. 
The only fit parameter is the dynamic DM interaction, $qD_Q=0.13$\,cm$^{-1}$. 
}
\label{EparCIntensities}
\end{figure}

The line areas measured in different geometries as a function of magnetic field strength 
are plotted in Fig.\,\ref{EparCIntensities}.
We used the method of differential absorption,
where spectra taken in different magnetic fields are subtracted from each other
to detect weak transitions.
In this differential method the transition 
from the singlet state to the $m_S=0$ triplet state escapes detection
(unless the intensity depends on the field) since the energy of this triplet 
level does not change with magnetic field.
Therefore, only the intensities of the transitions to the 
$m_S=-1$ and $m_S=1$ triplet levels can be detected and are plotted in Fig.\,\ref{EparCIntensities}.
An interesting finding is  the enhancement of the singlet to triplet transition 
in  $\mathbf{E}_1\!\parallel \!\mathbf{c}$ polarization close to the 68\,cm$^{-1}$ excitation.
It is natural to associate  the oscillator strength of this weak transition
with the interaction between the  spins and the 68\,cm$^{-1}$ excitation,
which has a dipole moment along the c-axis.
The exception is $\mathbf{B}_0\!\parallel \!\mathbf{b}$ orientation where there is no enhancement
in $\mathbf{E}_1\!\parallel \!\mathbf{c}$ polarization.
In this field direction only the  transition to the $m_S=0$ level
has a non-zero oscillator strength but is not detected because of the measurement method.

Besides the strong $\mathbf{E}_1 \parallel \mathbf{a}$ absorption and
the resonantly enhanced $\mathbf{E}_1 \parallel \mathbf{c}$ absorption
there is a field-independent oscillator strength, as is seen in Fig.\,\ref{EparCIntensities}(a).
No polarization anisotropy is observed there in contrary to the first two mechanisms 
of singlet to triplet transitions that can be recognized by  their polarization dependence.
In high fields the transitions
to the $m_S \! = \! -1$ and $m_S \! = \! 1$ levels have the same strength in 
$\mathbf{E}_1\!\parallel \!\mathbf{b}$ 
and $\mathbf{E}_1\!\parallel \!\mathbf{c}$ polarizations (Fig.\,\ref{EparCIntensities}(a)).
When $\mathbf{B}_0\!\parallel \!\mathbf{c}$ and $\mathbf{E}_1\!\parallel \!\mathbf{b}$ 
(Fig.\,\ref{EparCIntensities}(b)) the intensity of the 
transitions to the $m_S \! = \! -1$ and $m_S \! = \! 1$ levels is zero 
and in this configuration the transition to the $m_S \! = \! 0$ level is active.
More detailed analysis of this mechanism is given in \ref{MagDipAnal}
where we associate this with a magnetic dipole
transition.

Within the error limits the g-factors of the triplet state are the same
for the two in-plane field orientations $g_a=1.97\pm0.02$,
$g_b=1.975\pm0.004$. 
The fit of the resonance frequencies
of the singlet to triplet transition for the third field orientation,   
$\mathbf{B}_0\!\parallel \!\mathbf{c}$, 
gives   $g_c=1.90\pm 0.03$.

In zero field the triplet levels $m_S \! = \! -1, 0$, and  1  
are degenerate; it is best seen 
when comparing the zero field line positions
in Fig.\,\ref{EllAfieldSpectra4}(c) and (d).
Determining the size of the zero field splitting is limited by the linewidth. 
We can say that the zero field splitting of the triplet levels
is less than half of the linewidth, FWHM/2=0.25\,cm$^{-1}$.

\section{Discussion}

\subsection{ Triplet state and Dzyaloshinskii-Moriya interaction}

In this section we  analyze three different contributions 
to the singlet to triplet optical absorption observed experimentally in $\alpha'$-NaV$_2$O$_5$.
Two of them fall into the same category, the electric dipole active dynamic
DM mechanism. 
The third contribution is probably due to the static DM interaction mechanism where magnetic 
dipole transitions are active, but cannot be explained  with a single dimer model.

We rule out a possible singlet to triplet absorption mechanism based on a staggered g-factor. 
The staggered g-factor mechanism requires that 
the principal axes of the g-tensors of a pair of spins with anisotropic g-factors must not coincide.
Let us consider for an example the spin $\mathbf{S_1}$ with its principal axes 
rotated from the crystal a-axis by an angle $\theta$ and the spin $\mathbf{S_2}$ 
by an angle $-\theta$ and $\mathbf{B}_0 \parallel \mathbf{a}$.
Let the hamiltonian contain isotropic exchange interaction $J\mathbf{S}_1 \cdot \mathbf{S}_2$
and Zeeman interaction 
$\mathbf{B}_0 \cdot ( \mathbf{\tilde{g}}_1 \cdot \mathbf{S}_1 + \mathbf{\tilde{g}}_2  \cdot \mathbf{S}_2)$.
Matrix elements between the singlet and triplet states 
equal to $\pm B_0 (g_a-g_b)\sin\theta \cos\theta$ appear.
It is important  that the singlet-triplet mixing is proportional to $B_0$.
\textit{First}, there is no mixing in zero field ($B_0=0$) and the staggered g-factor 
mechanism is turned off.
\textit{Second}, the transition probabilities increase as $(B_0)^2$ if the magnetic field is small 
compared to the separation of the singlet and triplet states.
In the experiment we observe a singlet to triplet transition in zero field, and also the observed 
magnetic field dependence is different from that of the staggered g-factor mechanism. Therefore 
this mechanism does not apply to the singlet-triplet transitions in $\alpha'$-NaV$_2$O$_5$.

\subsubsection{
Dynamic Dzyaloshinskii - Moriya: 
$\mathbf{E}_1\!\parallel \!\mathbf{a}$}\label{E1par_a_discussion} 

The strongest singlet to triplet absorption is observed in 
$\mathbf{E}_1\!\parallel \!\mathbf{a}$ polarization.
There is no resonant enhancement in the magnetic field dependence of the
line area as seen in Fig.\,\ref{EparA_Bdep33T}
and any of the  optically active excitations in $\mathbf{E}_1\!\parallel \!\mathbf{a}$ 
polarization is a candidate that can create the dynamic DM interaction.
Nevertheless we can make some choices.
The 91\,cm$^{-1}$ a-axis phonon is not active since no enhancement 
is observed when the upper branch of 
the triplet resonance crosses the phonon frequency at 28\,T (Fig.\,\ref{EparA_Bdep33T})
although there is an interaction between 
this phonon and the continuum of magnetic excitations 
as is manifested by the Fano line shape of the phonon line above $T_c$
(Fig.\,\ref{Fig_E_ab_4_40}).

In Fig.\,\ref{EparA_Bdep33T} we have plotted two fit curves
based on  perturbation calculation results
(Eq.-s \ref{perp_perturbation_minus}, \ref{perp_perturbation_plus}).
In one case the phonon frequency was fixed to 518\,cm$^{-1}$
that is the strongest a-axis optical phonon\cite{Damascelli2000}. 
In the other case the phonon frequency  was a  fitting parameter giving us ($180\pm10$)\,cm$^{-1}$.
This value represents  the lowest boundary for the frequency of the  DM phonon.
Phonons with lower frequencies would give a too steep magnetic field dependence of the 
singlet to triplet transition probability.
It is likely that the lowest boundary of the phonon frequency has been underestimated
since the three high field data points above 30\,T influence  the fit
by lowering the phonon frequency.
Additional measurements above 33\,T are required to clarify this intensity enhancement.
Based on these fits the  strength of the dynamic DM interaction $qD_Q$ can be obtained.
The closest low $T$ phase infrared active a-axis phonon in frequency to 180\,cm$^{-1}$ is  
the 199\,cm$^{-1}$ zone folded phonon\cite{Damascelli2000}.
The plasma frequencies  $\Omega_p$ of  the 199\,cm$^{-1}$ and  
518\,cm$^{-1}$ phonons are 48\,cm$^{-1}$ (Ref.\cite{Popova2002}) 
and 853\,cm$^{-1}$ (Ref.\cite{SmirnovD1999}), respectively.
We convert  the plasma frequency    
into the line area in absorbance units, 
$ \int \alpha(\omega)\mathrm{d}\omega $, 
using 
$  \int \alpha(\omega)\mathrm{d}\omega =\pi^2\Omega_p^2 / n_a $.
From the  fit results we calculate  $qD_Q=5$\,cm$^{-1}$ for the 199\,cm$^{-1}$ phonon and
$qD_Q=0.9$\,cm$^{-1}$ for the 518\,cm$^{-1}$ phonon.

\begin{figure}[tb]
    \includegraphics[width=8.6cm]{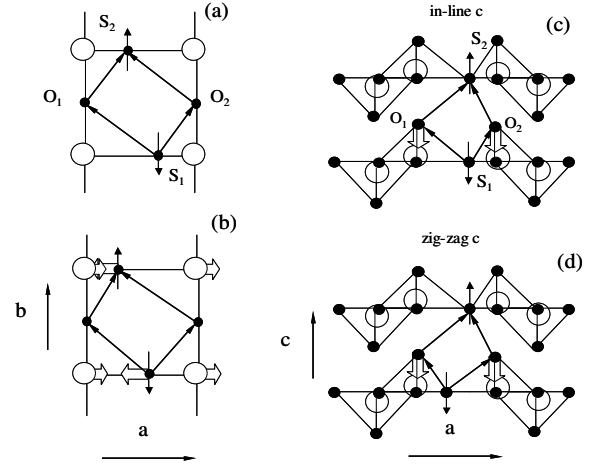}
\caption{Cartoons of superexchange paths in the ab-plane (panels (a) and (b))
with one ladder shown
and in the ac-plane ((c) and (d)) with two ladders shown.  
For illustrative purposes it is assumed that 
the spin is located on the rung oxygen. Oxygens  are shown by  filled circles
and vanadium atoms by open circles.
 Block arrows show the displacement of atoms due to a phonon.  
(a) superexchange paths between two spins over on-leg oxygens
within the same ladder in the zig-zag ordered phase; 
(b) same superexchange paths when an a-axis phonon is involved;
(c) superexchange paths between two spins in the neighboring planes over apical oxygens
displaced by a c-axis phonon
when spins are in-line along c-axis or (d) zig-zag.}
\label{Graph1}
\end{figure}

The orientation of the DM vector is determined by the vector product of the 
vectors connecting two spins over the superexchange path\cite{Keffer1962,Lohmann2000}
\begin{equation}
\frac{ \mathbf{D} }{| \mathbf{D} |} = [ \mathbf{ R}_{S_1,O_i}  \times \mathbf{ R}_{O_i,S_2}   ]
\label{DMvector}
\end{equation}
In the zigzag ordered low $T$ phase two superexchange paths exist, one over
on-leg oxygen O$_1$ and the other over O$_2$, shown in Fig.\ref{Graph1}(a).
The resultant DM vector is zero since the two DM vectors, pointing in $\mathbf{c}$-direction, 
cancel each other.
The zero length of the DM vector follows from the general arguments of symmetry too
as in this particular case there exists a local center of inversion located between  two V-O-V rungs.
The  518\,cm$^{-1}$ a-axis phonon displaces on-rung oxygens along the rung
(on-rung V-O-V stretching mode)\cite{Popovic2002}, Fig\ref{Graph1}(b).
Because of the phonon the two superexchange paths are not``equal" any more and
 the   resulting DM vector points along the c-axis.
The  orientation of the dynamic DM vector, $ \mathbf{D}_Q \parallel \mathbf{c} $,
is consistent with the selection rules for the dynamic DM interaction
observed  experimentally.
The transition to the $| \mathrm{t}_00 \rangle$ is observed when 
$ \mathbf{B}_0 \parallel \mathbf{D}_Q $, Fig.\,\ref{EllAfieldSpectra4}(d).
The transitions to the $|  \mathrm{t}_- 0\rangle$ and $|  \mathrm{t}_+ 0\rangle$ 
states are observed when 
$ \mathbf{B}_0 \perp \mathbf{D}_Q $, Fig.\,\ref{EllAfieldSpectra4}(a), (b), (c).

There is another phonon with an oscillator strength  larger than that of
the 518\,cm$^{-1}$ a-axis phonon. 
That is the 582\,cm$^{-1}$ b-axis phonon that stretches on-leg 
V-O-V bonds\cite{SmirnovD1999,Damascelli2000}.
In the high T phase where electron charge (and spin)
is rung-centered, such distortion does not produce any 
DM interaction from the principles of symmetry.
In the low T phase the charges are ordered in a zigzag  pattern. 
The  on-leg V-O-V bond stretching phonon lowers the 
symmetry and creates the DM interaction  along the c-axis.
We do not observe a singlet to triplet absorption with 
an electric dipole moment along the b-axis in the experiment.
There are two possibilities why the $\mathbf{E}_1 \parallel \mathbf{b}$ 
absorption is not observed.
\textit{First}, there is no zigzag charge order.
\textit{Second}, there is a zigzag charge order and 
although the dynamic DM is allowed by symmetry,
the actual value of  $qD_Q$ is  small and 
the absorption cannot be observed  in our experiment.
Since there is a mounting evidence in the favor of a low T 
zigzag charge order\cite{Nakao2000,Grenier2002,Smaalen2002}  we consider the second case likely.

\subsubsection{\label{caxiseldip}
Dynamic Dzyaloshinskii - Moriya: $\mathbf{E}_1\!\parallel \!\mathbf{c}$}

The singlet to triplet absorption with the electric dipole moment along the 
c-axis  is due to the dynamic DM interaction.
In this particular case the dynamic DM mechanism is brought in 
by the 68\,cm$^{-1}$ optically active c-axis  phonon.
The enhancement of the singlet-triplet absorption close to the 68\,cm$^{-1}$ line
is present  if the magnetic field is either parallel to a- or b-axis, but missing
if the field is along the c-axis. 
In Section~\ref{phonons} we  present further arguments supporting 
the assignment of the 68\,cm$^{-1}$ resonance to a phonon and not to 
a magnetic excitation.

In Fig.\,\ref{EparCIntensities} the fit of the singlet to triplet transition 
intensities to the dynamic DM absorption mechanism is shown.
The input parameters are the resonance frequency of the phonon, 
$\omega_p=68$\,cm$^{-1}$, and the frequency of the singlet to triplet transition
as the function of magnetic field, $\omega_{T_{\pm}}= \Delta \pm g\mu _B B_0$,
where $ \Delta = 65.4$\,cm$^{-1}$.
We estimated the low temperature oscillator strength of the phonon
from the $T$ dependence presented in Fig.\,\ref{Spectra68_Oct_fig}
and got  $\Omega_p^2 = 400$\,cm$^{-2}$.
The ratio of the singlet to triplet absorption oscillator strength $ \Omega_{ST}^2 $
to the oscillator strength of the phonon, $\Omega_p^2$, is 
$ \Omega_{ST}^2 / \Omega_p^2 \sim 0.13/400=3.3\times 10^{-4}$.
The only fit parameter is the strength of the dynamic DM interaction,
$qD_Q=0.13$\,cm$^{-1}$.

We used $\mathbf{D}_Q \! \parallel \! \mathbf{b}$ in our fit because
the splitting of the triplet in the magnetic field is observed if
$\mathbf{B}_0 \! \parallel \! \mathbf{a}$ or 
$\mathbf{B}_0 \! \parallel \! \mathbf{c}$ (Fig.\,\ref{EparCIntensities}).
According to the selection rules for the dynamic DM only the transitions to the triplet 
states with $m_S=1$ and $m_S=-1$ are observed 
when the magnetic field is perpendicular to the dynamic DM vector 
$\mathbf{D}_Q \! \perp \! \mathbf{B}_0$.
The selection rules for the dynamic DM if 
$\mathbf{B}_0 \! \parallel \! \mathbf{D}_Q \! \parallel \! \mathbf{b}$ 
allow only transitions to the $m_S=0$ level 
that does not shift with the magnetic field and we do not see it
in the differential absorption spectra that are taken in different magnetic fields.
The background intensities that do not depend on the magnetic field in 
figure \ref{EparCIntensities}(a) are analyzed in the next section (\ref{MagDipAnal}).

Which lattice deformations along the c-axis (electric dipole!) will give
the dynamic  DM interaction  in the b-axis direction? It turns out that we have to consider 
inter-plane interactions.
For the beginning let us consider two spins on the neighboring rungs, 
as shown in Fig.\ref{Graph1}(a), where 
optical c-axis phonons have the out of (ab) plane anti phase movements of oxygen and 
vanadium atoms.
In the low $T$ zigzag ordered phase 
the dynamic DM vector will have  components along both, a- and b-axis.
The b-axis component of the dynamic DM turns to zero for a vanishing zigzag order,
i.e. in the limit of rung-centered spin distribution. 
The a-axis component stays nonzero.
To get a dynamic DM exclusively along the b-axis
we have to consider interactions between the spins in the neighboring planes.
Here the path for the DM interaction between spins in the neighboring   planes
goes  over the apical oxygens.
Two arrangements along the c-axis are possible  in the zigzag ordered phase, 
shown in Fig.\ref{Graph1}(c),(d): in-line or zigzag.
In both arrangements the displacement of the apical oxygens in the $\mathbf{c}$-direction will create 
a dynamic DM along the b-axis.
As one can see not only the dynamic, but also the static DM in the $\mathbf{b}$-direction
is allowed by the symmetry.

In the experiment we do not see magnetic dipole active optical transitions
caused by the static DM interaction.
To compare the strength of magnetic and electric dipole transitions
we must know the magnitude of the static DM interaction.
The magnitude of the DM interaction was estimated by Moriya\cite{Moriya1960},
$D\approx (|g-g_e|/g_e)J$,
where  $J$ is isotropic exchange interaction and $g_e=2$ is  
the free electron g-factor.
We use $g=1.90$ and $J=60$\,meV (from Ref.\cite{Grenier2001}) and get $D=25$\,cm$^{-1}$. 
The transition intensity $I$ is proportional to $(|V|/ \delta \mathcal{E} )^2$ 
where $|V|$ is the matrix element of the interaction
($D$ or $qD_Q$) between the two states 
and $\delta \mathcal{E}$ is their energy separation.
For the static DM effect $\delta\mathcal{E}$ is the singlet-triplet gap, 
$\delta\mathcal{E}=65.4$\,cm$^{-1}$;
for the dynamic DM it is the energy difference
between the phonon energy and the triplet level,
$\delta\mathcal{E}=68-65.4= 2.6$\,cm$^{-1}$.
The intensity of the magnetic dipole transition is weaker 
than the electric dipole transition by the factor of
$\alpha_f^{-2}  = 137\,^{2}$, where
$\alpha_f$ is the fine structure constant (see Ref.\cite{Loudon1983} p.171).
If we use $D=25$\,cm$^{-1}$  and $qD_Q=0.13$\,cm$^{-1}$  
we get that the intensity due to the dynamic  mechanism is 
$( 0.13\times 65.4)^2/(\alpha_f 25\times 2.6)^2 \approx300$ times
larger than the intensity due to the static mechanism. 
Optical transitions due to the static DM interaction are suppressed
with respect to the transitions caused by the dynamic DM  
because electric dipole transitions are stronger than magnetic dipole transitions.

Our conclusion is that the dynamic DM interaction 
along the b-axis, $qD_Q=0.13$\,cm$^{-1}$, is between the spins on the ladders 
of the neighboring planes. 
The 68\,cm$^{-1}$ c-axis phonon that creates the dynamic DM interaction
involves the displacement of apical oxygens.

\subsubsection{\label{MagDipAnal}Third optical singlet-triplet absorption mechanism}

There is a third mechanism for the optical triplet absorption
that is responsible for the magnetic field independent intensities of the transitions to the
$m_S\!=\!-1$ and $m_S\!=\!1$ triplet levels,
shown in Fig.\,\ref{EparCIntensities}(a) by solid triangles and empty circles.
The same mechanism contributes together with the enhanced part discussed in the previous section
to the intensity plotted with empty triangles on the same graph.

We could assume that  the third mechanism is also an  electric dipole absorption mechanism
as the two mechanisms ascribed to the singlet-triplet absorption in 
$\mathbf{E}_1\parallel \mathbf{a}$ and $\mathbf{E}_1\parallel \mathbf{c}$
polarizations.
Since this  absorption is present in both polarizations, 
$\mathbf{E}_1\parallel \mathbf{b}$ and $\mathbf{E}_1\parallel \mathbf{c}$
 one has to assume that the electric dipole moment 
is either in the (bc)-plane or, just by coincidence, two electric dipole mechanisms,
one polarized along the b-axis and the other polarized along the c-axis,
give the same intensities.
The case that the optical phonon responsible for the dynamic DM effect
has a dipole moment in the (bc)-plane contradicts 
with the data available on phonons and with the crystal symmetry.
The second case of coinciding intensities is ruled out 
by the selection rules  if applied to the full data set presented 
in Fig.\,\ref{EparCIntensities}(a) and (b).

The third mechanism could be a magnetic dipole DM mechanism.
By applying the selection rules to  the data we should be able  
to determine the orientation of the DM vector.
According to the theory (section \ref{MagDipTheory}) transitions to the $m_S\!=\!-1$ and $m_S\!=\!1$
states have constant and equal intensities 
when $\mathbf{B}_0 \! \parallel \! \mathbf{D}$. 
This condition is satisfied by two data sets, 
solid triangles ($\mathbf{H}_1 \! \parallel \! \mathbf{c}$) and 
empty triangles ($\mathbf{H}_1 \! \parallel \! \mathbf{b}$)
in  Fig.\,\ref{EparCIntensities}(a), measured
with $\mathbf{B}_0 \! \parallel \! \mathbf{a}$. We have $\mathbf{D}\parallel \mathbf{a}$, where
$I_x^{\pm}=I_y^{\pm}$ and $x \equiv b$ and $y \equiv c$ (see Fig.\,\ref{MagDipHpar}).
However the set represented by the circles in Fig.\,\ref{EparCIntensities}(a)
is the $I_y^{\pm}$ intensity in the perpendicular 
configuration, $\mathbf{B}_0\perp\mathbf{D}$ (in our notation 
$\mathbf{D}\parallel \mathbf{y} \parallel\mathbf{a}$), which 
should have zero intensity according to the theory (see Fig.\,\ref{MagDipHperp}).
Also, in the $\mathbf{B}_0\parallel \mathbf{H}_1\parallel \mathbf{c}$ 
configuration $I_z^{\pm}$ should have non-zero intensities.
In the experiment, filled triangles in 
Fig.\,\ref{EparCIntensities}(b), no intensity is observed contrary to the theory.
Therefore the assumption $\mathbf{D}\parallel \mathbf{a}$ 
is not consistent with the full data set.
Also, we can prove that neither $\mathbf{D}\parallel \mathbf{b}$ 
nor $\mathbf{D}\parallel \mathbf{c}$ is fully consistent with the experiment.

Our conclusion is that the third optical triplet absorption mechanism is a 
magnetic dipole transition, but cannot be explained by an isolated dimer model 
with DM interactions.
Apparently, a more elaborate model including the inter-dimer DM interactions, is necessary.

\subsection{Spin gap and  phase transition}

We studied the effect of temperature and magnetic field on the 
singlet to triplet transition.
The results are shown with circles in Fig.\,\ref{Tdep65_68}.
We observe that the intensity of the singlet to triplet transition (Fig.\,\ref{Tdep65_68}(a))
follows the intensity of the X-ray diffraction peak\cite{Gaulin2000} reported in the same figure.
Gaulin \textit{et al.} have shown that the intensity of the X-ray diffraction peak can be fitted with 
a single function  $I=t^{\, \beta}$
over the reduced temperature, $t=1-T/T_c$, in the range from $6\times10^{-3}$
to $2\times10^{-2}$ with  the critical exponent $\beta=0.18$. 
Our data taken  below 30\,K  is above $1.1\times10^{-1}$ in the 
reduced temperature scale, 
which is unsuitable for the determination of the critical exponent.

The singlet-triplet splitting (spin gap), shown in Fig.\,\ref{Tdep65_68}(b), 
is more rigid than the singlet to triplet transition probability
or the X-ray scattering intensity.
The rigidity of the spin gap has been confirmed earlier 
by INS measurements\cite{Fujii1997} and 
by high field electron spin resonance measurements, 
although not in the zero magnetic field\cite{Nojiri2000}. 
The far-infrared S to T gap is in  agreement with the INS gap
also shown in Fig.\,\ref{Tdep65_68}(b).
In the ultrasonic experiment\cite{Fertey1998}  the measurements were extended 
close enough  to $T_c$
that one is able to determine the critical exponent for the spin gap.
The temperature dependence of the spin gap, measured indirectly by the ultrasonic probe, 
gives the critical exponent $\beta=0.34$ below the reduced temperature $10^{-2}$.
Although far-infrared measurements of the gap are direct,
reasonable data can be obtained only too far from $T_c$ as the absorption 
line broadens and gets weak
and therefore  the fit, inset to Fig.\,\ref{Tdep65_68}(b),
gives us an exceptionally low critical exponent, $\beta=0.039\pm 0.002$.

\begin{figure}[tb]
\includegraphics[width=8.6cm]{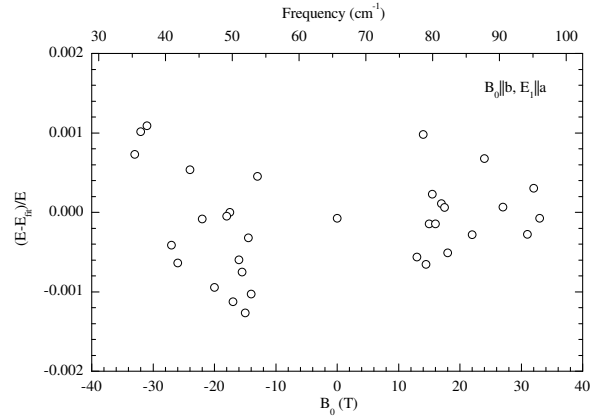} 
\caption{Relative deviation $(E-E_{fit})/E$
of S to T$_{-1}$ (plotted at negative field values)
and S to T$_{+1}$ (plotted at positive field values) transition energy $E$
from the linear fit $E_{fit}$.
}
\label{EparA_BdepGfac}
\end{figure}

The magnetic field dependence of the spin gap and 
the triplet state g-factor have been measured up to 33\,T
for one field orientation, $\mathbf{B}_0\parallel \mathbf{b}$, at 4.7\,K.
The line positions of the transitions from the singlet ground state
to the triplet states with $m_S \! = \! \pm 1$ 
were fitted with a linear function $E_{fit} = \Delta \pm g\mu _B B_0$.
The reduced residual of the linear fit is less than 0.0012 of the transition energy, 
as shown in  Fig.\,\ref{EparA_BdepGfac}.
This means that the triplet state g-factor is not renormalized by the magnetic field as high as 33\,T
and the spin gap is independent of the field up to 33\,T.
The independence of the gap value $\Delta$ on the field at low temperature
 is not surprising\cite{Azzouz1996}.
The role of the magnetic field is to couple to thermally excited quasiparticles. 
At low temperature the number of quasiparticles is small and hence the effect of the magnetic field
is negligible.

In conclusion, the temperature dependence of the intensity of the  singlet to triplet transition agrees 
with the X-ray scattering intensity temperature dependence.
Also, the singlet-triplet splitting at 8.13\,meV has the same temperature dependence as
the singlet-triplet splitting of the second excitation branch at 9.8\,meV measured by 
the INS.
The 8.13\,meV spin gap is not altered by the magnetic field at least  up to  33\,T.

\subsection{\label{phonons}Phonons and  phase transition}

\subsubsection{\label{c-phonons} c-axis phonons}

In Section~\ref{caxiseldip} 
while calculating the dynamic DM interaction for $\mathbf{E}_1\!\parallel \!\mathbf{c}$
we assumed that the electric dipole moment 
of the c-axis polarized optical singlet to triplet transition comes from the interaction between 
the spin system and the 68\,cm$^{-1}$ optical c-axis phonon.
An alternative would be that the 68\,cm$^{-1}$ resonance is not a phonon
but a singlet electronic excitation.
Here we  analyze  existing data and show that the data  is not in contradiction 
with the assumption that the  68\,cm$^{-1}$ resonance is a phonon mode.

\begin{figure}[tb]
    \includegraphics[width=8.6cm]{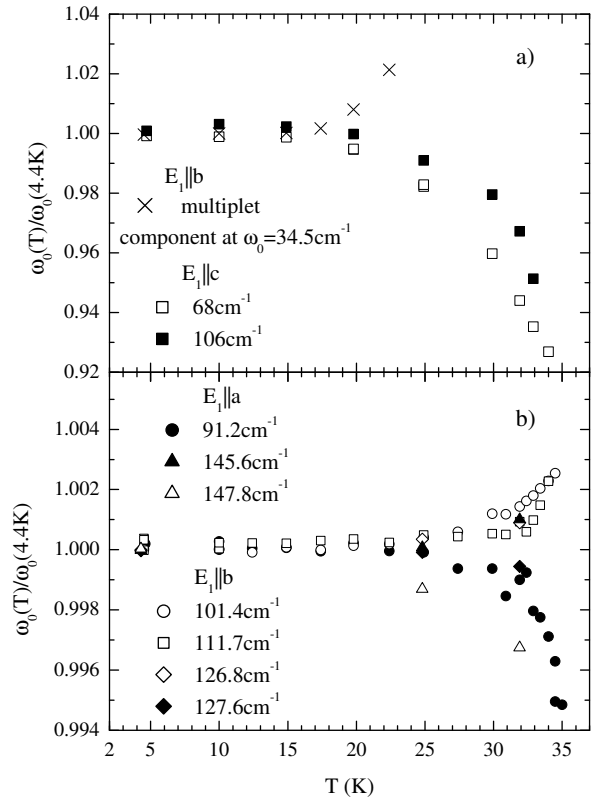}
\caption{Temperature dependence of normalized phonon frequencies.
(a) - b-axis multiplet and c-axis phonons, (b) - a- and b-axis phonons.  
}
\label{FreqTdepNorm}
\end{figure}

There is a series of infrared-active c-axis modes appearing below 34K 
as shown in Fig.\,\ref{figEparC4and40} and found by other groups\cite{Popova2002,Takehana2001}.
The two low frequency modes, at 68 and 106\,cm$^{-1}$, are the strongest.
The temperature dependence of their resonance frequencies plotted
in Fig.\,\ref{FreqTdepNorm}(a) is an order of magnitude larger
as compared to the a- and b-polarized modes plotted in Fig.\,\ref{FreqTdepNorm}(b).
Modes at  frequencies close to 68 and 106\,cm$^{-1}$ have been found by Raman 
spectroscopy\cite{Lemmens1998,Konstantinovic2001}.
Sample dependence of mode frequencies by few wavenumbers
 has been observed both in infrared\cite{Popova2002} 
and Raman\cite{Konstantinovic2001}.
It is known that Na deficiency of $\alpha'$-Na$_{x}$V$_2$O$_5$
 affects $T_c$ and in samples 
where  $x=0.97$ the phase transition is suppressed completely\cite{Isobe1998}.
A Na doping dependence study\cite{Konstantinovic2001} has shown that the frequency 
variation of Raman modes from sample to sample is related to the Na content of the sample.
Raman measurements\cite{Blumberg2001} on the sample from the same batch as the one studied here 
gave values for the mode frequencies 65.9 and 105\,cm$^{-1}$ at 4.4\,K.
This means that Raman and infrared spectroscopy observe different modes
since the frequencies of infrared modes are 68 and 106\,cm$^{-1}$. 
Neither of the Raman modes\cite{Lemmens1998,Konstantinovic2001} nor the infrared modes (this 
work and Ref.\cite{Takehana2001}) split in the magnetic field.
The comparison of the $T$ dependence of the normalized mode frequency has shown that 
the $T$ dependence of the 65.9\,cm$^{-1}$ Raman mode follows the $T$ dependence of the 68\,cm$^{-1}$
infrared mode and the 105\,cm$^{-1}$ Raman mode follows the $T$ dependence of the 106\,cm$^{-1}$
infrared mode.
Therefore, although the  infrared and the Raman modes do not have the same frequencies,
it is likely that the origin of the infrared and Raman modes near 68  and 106\,cm$^{-1}$ is the same, 
lattice modes   or electronic excitations.

The origin of the low frequency infrared and Raman modes has been under the debate.
It has been concluded that the Raman modes are not lattice 
vibrations\cite{Lemmens1998,Konstantinovic2001}.
In Ref.\cite{Popova2002} the conclusion 
was that the infrared modes at 68 and 106\,cm$^{-1}$  are zone-folded c-axis phonons and  
it was speculated that their peculiar $T$ dependence compared to other zone-folded phonon modes
is caused by the interaction of phonons with charge and spin degrees of freedom. 
Indeed, the 68\,cm$^{-1}$ mode and the singlet to triplet excitation at 65.4\,cm$^{-1}$
have similar $T$ dependencies of their infrared absorption line parameters 
as shown in Fig.\,\ref{Tdep65_68}.
Assuming that 68 and 106\,cm$^{-1}$ excitations are zone folded phonons we would have to explain
why their frequencies depend on Na deficiency
and why their frequencies have a different temperature 
dependence than the other zone-folded modes have.
We argue that both points can be explained if charge correlations 
develop inside ladder planes prior to the transition to the low $T$ phase 
and at the phase transition  three-dimensional correlations build up between the planes.

Charge ordering within ladder planes precedes the lattice distortion 
and the opening of the spin gap as evidenced by the vanadium Knight shift and the
sodium quadrupolar and Knight shifts\cite{Revurat2000}.
The presence of two-dimensional charge correlations in  the ladder planes
 above 35\,K is also supported by the  X-ray diffraction measurements\cite{Gaulin2000}.
In $\alpha'$-NaV$_2$O$_5$ there are   modulated ladders 
with a zigzag charge order within one plane\cite{Ohama1999,Revurat2000,Nakao2000}.
It is possible to construct four different planes with the zigzag charge order
and when stacked in certain sequence along the c-axis a unit cell is  formed 
compatible with the observed X-ray structure\cite{Smaalen2002,Grenier2002}. 
By applying pressure the critical temperature is reduced and it has been found that the 
soft axis is the c-axis\cite{Loa1999,Kremer1999Loa}.
A ``devil's staircase" like sequence of phase transitions 
to the phases with  unit cells incorporating more than four planes  
in the   $\mathbf{c}$-direction takes place under pressure larger than 0.5\,GPa\cite{Ohwada2001}.
Therefore the interactions between layers are important in the formation of three
dimensional correlations.

Interactions between ladder planes are important for those modes,
which are zone-folded along the c-axis. 
The primary candidates are the modes with anti-phase movements of atoms along the c-axis 
in the neighboring planes, which are naturally the modes 
that are polarized along the c-axis.
The c-axis modes have the largest
frequency shift close to the phase transition point.
The exception is the b-polarized multiplet (Fig.\,\ref{Multiplet2}).
It  is the only non-c-axis low frequency mode, which
has a  relative change of  frequency 
with temperature (Fig.\,\ref{FreqTdepNorm}(a)) that is comparable to the c-axis modes.
Considering the low frequency of the multiplet it is likely that the seven lines are 
a result of the folding of a b-polarized acoustical, not optical phonon branch.
To get seven optical  phonon branches out of one phonon branch 
a c-axis folding is required in addition to a- and b-axis foldings.
Therefore this multiplet involves relative movements of atoms in neighboring
layers and is a subject to interlayer couplings.

Our conclusion is that the c-axis phonon modes are most susceptible
to three dimensional correlations between ladder planes
and therefore their $T$ dependence is different from other low frequency
phonon modes.
We assign the 68\,cm$^{-1}$ (and 106\,cm$^{-1}$) infrared active mode 
and the 66\,cm$^{-1}$ (and 105\,cm$^{-1}$) Raman active modes to the zone-folded phonons.

\subsubsection{Polarization dependence of line intensities}

In NaV$_2$O$_5$ the a- and b-axes are not equivalent and
one would expect the selection rules
to apply so that the phonon active in the
$\mathbf{E}_1\!\parallel \!\mathbf{a}$
absorption is missing in $\mathbf{E}_1\!\parallel \!\mathbf{b}$
spectrum and vice versa. 
In other words, phonons polarized along a-axis should have different frequencies
than phonons  polarized along the b-axis.

\begin{table}[b]
\caption{Rotation angle $\varphi $  calculated from the two dipole model.
$\varphi _a$ ($\varphi _b$)
is the rotation angle of the phonon dipole moment from the  crystal a-axis (b-axis).}

\begin{tabular}{l |ddddd}

$\omega_0$(\,cm$^{-1}$)   & 101.4 & 101.7     & 111.7 & 126.8     & 127.5 \\
 \hline
$\varphi _b \equiv  \varphi$       & -        & -            & 32       & -        & 35 \\
$\varphi _a = 90^{\circ} - \varphi $                              & 37       & 0            &   -      & 31       & -    
\end{tabular}
\label{AngleTable}
\end{table}

This rule applies well at 40\,K (Fig.~\ref{Fig_E_ab_4_40}).
The strong 180\,cm$^{-1}$ phonon is missing in the
$\mathbf{E}_1\!\parallel \!\mathbf{a}$ spectrum
and the structure at 140\,cm$^{-1}$ is missing in the
$\mathbf{E}_1\!\parallel \!\mathbf{b}$ spectrum.
The 90.7\,cm$^{-1}$ line is present only in the
$\mathbf{E}_1\!\parallel \!\mathbf{a}$ spectrum.

\begin{figure}[tb]
    \includegraphics[width=8.6cm]{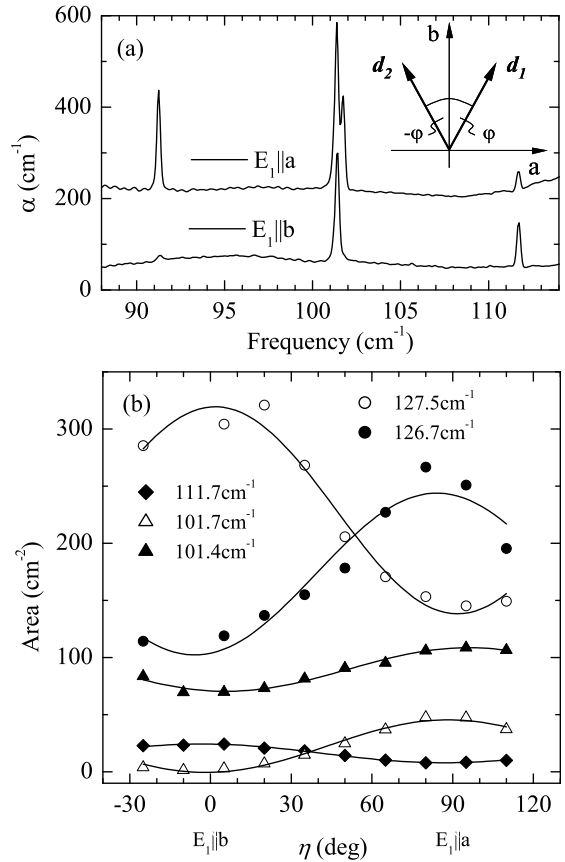}
\caption{Polarization dependence of low frequency phonon absorption  line intensities.
(a) - absorption spectra in $\mathbf{E}_1\!\parallel  \!\mathbf{a}$ 
  (shifted upwards by 150\,cm$^{-1}$) and $\mathbf{E}_1\!\parallel  \!\mathbf{b}$ polarizations.
(b) - absorption line areas as a function of polarization angle in the (ab)-plane; 
$\eta=0$ corresponds to $\mathbf{E}_1\!\parallel  \!\mathbf{b}$.
Inset to (a) shows the orientation of  electric dipoles rotated from the b-axis by an angle
$\varphi $ and $-\varphi $.
}
\label{FigPolDep}
\end{figure}

At 4.4\,K this selection rule does not apply to all  absorption lines.
The 101.4 and 111.7,  126.7 and 127.5\,cm$^{-1}$ phonons
(Fig.-s~\ref{FigPolDep}(a), \ref{Fig_E_ab_4_40})  have absorption  in both polarizations.
The leakage of the wrong polarization through the polarizer is ruled out.
For example, the 101.7\,cm$^{-1}$ line present in $\mathbf{E}_1\!\parallel \!\mathbf{a}$
spectrum is missing in $\mathbf{E}_1\!\parallel \!\mathbf{b}$
spectrum (Fig.~\ref{FigPolDep}a) and the multiplet
at 30\,cm$^{-1}$ is missing in $\mathbf{E}_1\!\parallel \!\mathbf{a}$
spectrum (Fig.~\ref{Fig_E_ab_4_40}).

We measured  the polarization dependence of line intensities in the (ab)-plane.
The areas of   absorption lines between 100 and 130\,cm$^{-1}$
are plotted in Fig.~\ref{FigPolDep}(b) as a function of the angle $\eta$
between the polarizer and the b-axis.
The lines at 91\,cm$^{-1}$  were not included since they are two different lines,
 one at 91.2\,cm$^{-1}$
in the a-polarization and the other at 91.3\,cm$^{-1}$ in the b-polarization.
As the 101.7\,cm$^{-1}$ line has zero intensity at 
$\eta = 0^\circ$, $\mathbf{E}_1\!\parallel \!\mathbf{b}$ it belongs to the phonon
modes representing the averaged crystal symmetry.
It is important  that the intensity of the
101.4\,cm$^{-1}$, 111.7\,cm$^{-1}$,  126.7\,cm$^{-1}$ and  127.5\,cm$^{-1}$
lines never falls to zero between
$\eta = 0^\circ$ and $90^\circ$.
A similar effect was found by Damascelli \textit{ et al.}\cite{Damascelli2000}
for several infrared active phonons at higher frequencies.
The simplest explanation would be that the zone-folded a- and b-polarized 
phonons are pair wise degenerate, but
then it is hard to explain why there is a constant intensity ratio 
of at least four pairs of low frequency and of one high frequency  (718\,cm$^{-1}$)
pair of a- and b-axis phonons, as is shown below.

We assume, as was done in Ref.\cite{Damascelli2000},
that   there are two regions in the crystal with  a symmetry different 
from the averaged crystal symmetry.
In one  region the electric dipole moment of a phonon, $\mathbf{d}$, is rotated from
the crystal  b-axis  by an angle $\varphi  $ and in the other region by $-\varphi  $
as shown in the inset to Fig.\ref{FigPolDep}(a).
The optical conductivity   in the first region is 
$\sigma_1(\varphi )=aE_1^2d^2\cos^2 (\varphi -\eta)$, where $a$ is a constant independent of angular 
parameters. 
$\eta$ is the angle between the b-axis and the electric field vector of
light $\mathbf{E}_1$.
The conductivity in the second region is
$\sigma_1(-\varphi )=aE_1^2d^2\cos^2 (\varphi +\eta)$.
Since the phonons are independent in the two regions the total observed 
conductivity is  
$\sigma_1 = \sigma_1(\varphi ) + \sigma_1(-\varphi ) = I_b\cos^2\eta+I_a\sin^2\eta$,
where 
$I_b=2aE_1^2d^2\cos^2\varphi $ and $I_a=2aE_1^2d^2\sin^2\varphi $ 
are the oscillator strengths     observed in the experiment 
in the b- and a-polarizations  respectively.
The solid lines in  Fig.\,\ref{FigPolDep}(b) are the fits of the integrated 
absorption line areas where  Eq.\,\ref{eq:alpha} has been used to convert 
$\alpha(\omega)$ into $\sigma_1(\omega)$.
The results are given in Table~\ref{AngleTable}.
Note that two of the absorption lines  belong to the dipoles rotated from the a-axis
and two to the dipoles rotated by approximately the same angle from the b-axis.
The averaged value  of  the rotation of the dipole moment is
$ \bar {\varphi}  = 34^\circ \! \pm \! 3^\circ$.
A similar result,  $\varphi  = 39^\circ$, we obtain for the 718\,cm$^{-1}$ phonon
using the data from the Ref.\cite{Damascelli2000}
where  the given oscillator strength in the $\mathbf{E}_1\!\parallel \!\mathbf{a}$ 
polarization is 0.021 and in   the $\mathbf{E}_1\!\parallel \!\mathbf{b}$ polarization is 0.014.

The two regions with a symmetry that is different from the crystal symmetry 
can be associated with two types of ladder planes with a zigzag charge order.
In Ref.\cite{Damascelli2000} it was assumed that the crystal is split into domains
with two different diagonal charge patterns.
Considering the recent X-ray studies\cite{Smaalen2002,Grenier2002} it is more likely
that there are no domains and two types of planes exist with 
different diagonal charge pattern instead.
The two different charge configurations come from the way
the two neighboring zigzag charge ordered ladders are 
positioned with respect to each other within the plane.
A shift of every second ladder by one half of the superlattice constant 
in the $\mathbf{b}$-direction 
creates two different structures where 
charged stripes of V$^{+4.5-\delta_c/2}$ run diagonal from left to the right or 
from right to the left in the (ab)-plane\cite{Smaalen2002}.
This diagonal charge order within a single ladder plane determines
the orientation of the phonon dipole moments.
It is possible to estimate the charge offset $x=l(1+\delta_c)/2$ from the center of the rung
knowing the tilt angle of the dipole moments $\varphi$.
Here $l$ is the length of the rung and the charge transfer factor
$\delta_c$ is associated with the formal valence of vanadium ions in the low $T$
zigzag ordered phase, V$^{+4.5-\delta_c/2}$ and V$^{+4.5+\delta_c/2}$.
The charge transfer factor depends only on the angle $\varphi$,
$\delta_c\approx \tan \varphi $,
since the rung length is approximately  equal to the distance between the nearest
neighbor V atoms along the leg.
We get  $\delta_c=0.67\pm0.07$ using $\varphi = \bar{\varphi} $.
From the analysis of INS data the authors of Ref.\cite{Grenier2001}
come to  a similar value, $\delta_c=0.6$.

Our study of phonon modes in the low temperature phase supports 
the view that the symmetry of the individual ladder planes is lower than the 
averaged crystal symmetry.
In addition it has been found by sound velocity measurements\cite{Schwenk1999}  that the $c_{66}$
shear mode couples to the pre-transitional charge fluctuations of $B_{1g}$ symmetry,
which correspond to the static zigzag charge order in the low $T$ phase.
There are phonon modes in  $\alpha'$-NaV$_2$O$_5$ in the low $T$ phase 
 where the normal coordinates are confined into single ladder planes.
These modes  show  the symmetry of an individual plane determined by the zigzag charge ordering. 
For other modes the  movement of atoms is correlated between the neighboring planes
or they are insensitive to the zigzag charge order 
and therefore they reflect the averaged crystal symmetry. 

From the analysis of the infrared spectra we conclude that two types of ladder planes exist
with the zigzag charge order where the charged stripes are aligned approximately 
in $[110]$ or in $[\bar110]$ directions. 
The formal charge of the vanadium atoms forming the zigzag pattern 
is $+4.17\pm0.04$ and $+4.83\pm0.04$.

\subsection{Continuum of excitations and Fano resonances}

\begin{figure}[tb]
    \includegraphics[width=8.6cm]{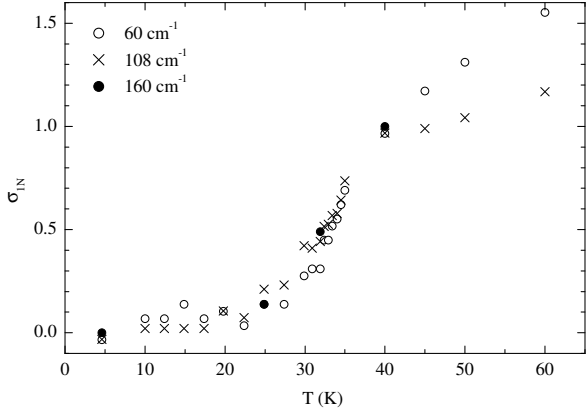}
\caption{Temperature dependence of the normalized optical conductivity 
in $\mathbf{E}_1 \! \parallel \! \mathbf{a}$ polarization
$\sigma_{1N}=[\sigma_{1a}(T)-\sigma_{1a}(4.4K)][\sigma_{1a}(40K)-\sigma_{1a}(4.4K)]^{-1}$
at 60 (open circles), 108 (crosses), and 160\,cm$^{-1}$ (filled circles).
}
\label{gap}
\end{figure}

\begin{figure}[tb]
    \includegraphics[width=8.6cm]{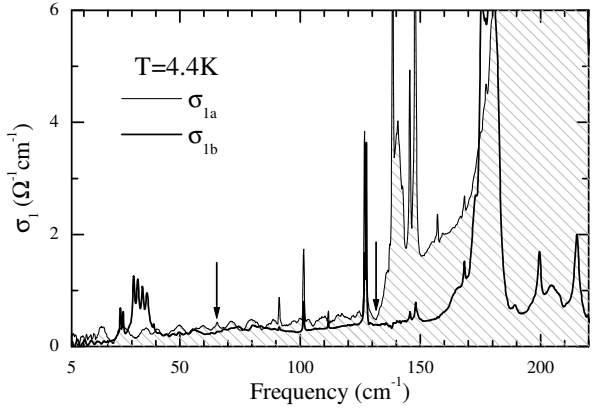}
\caption{Real parts of optical conductivities $\sigma_{1a}$ (shaded and thin line) in 
$\mathbf{E}_1\!\parallel  \!\mathbf{a}$ polarization and 
$\sigma_{1b}$ (thick line, $\mathbf{E}_1\!\parallel  \!\mathbf{b}$) at 4.4\,K. 
Arrows point to the singlet to triplet resonance at 65.4\,cm$^{-1}$ and to 
the onset of the  absorption continuum at 132\,cm$^{-1}$
in $\mathbf{E}_1\!\parallel  \!\mathbf{a}$ polarization.
}
\label{ab_4KGap}
\end{figure}

\begin{figure}[tb]
    \includegraphics[width=7.6cm]{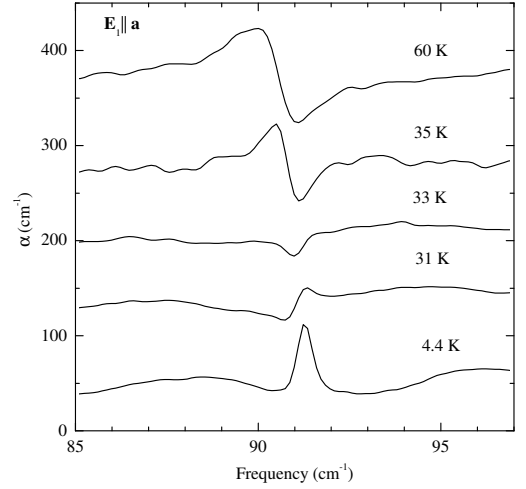}
\caption{Temperature dependence of the 91\,cm$^{-1}$ phonon 
absorption line 
in $\mathbf{E}_1\!\parallel  \!\mathbf{a}$ polarization.
Spectra have been off-set by 50\,cm$^{-1}$ in vertical direction.
}
\label{FanoWater}
\end{figure}

\begin{figure}[tb]
    \includegraphics[width=8.6cm]{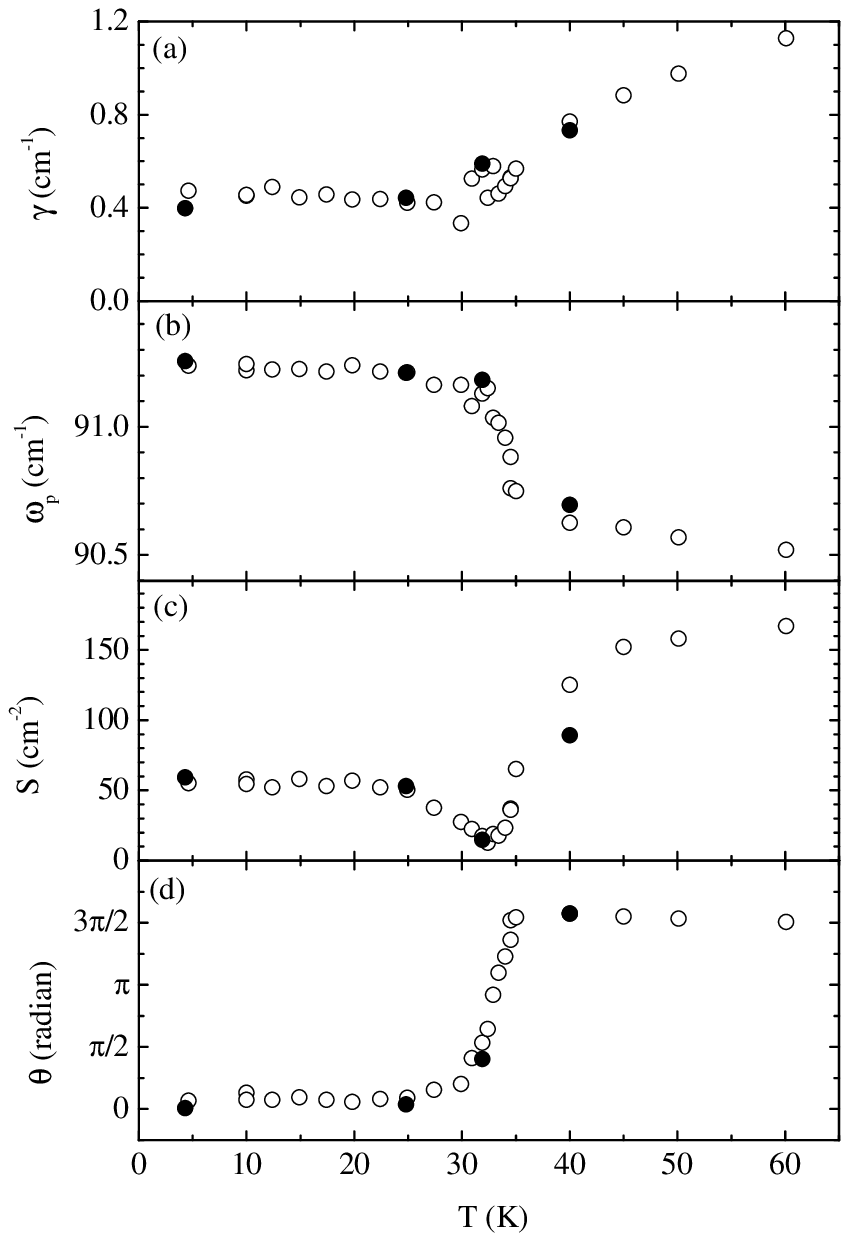}
\caption{Temperature dependence of the Fano fit parameters
of the 91\,cm$^{-1}$ phonon line: 
(a) - full width at half maximum $\gamma$,
(b) - phonon resonance frequency $\omega_p$, 
(c) - integrated line area $S$, 
(d) - asymmetry parameter $\theta$.
Open circles - 120\,$\mu$m thick sample; filled  circles - 40\,$\mu$m thick sample.
}
\label{Fanofit}
\end{figure}

A broad absorption continuum is observed in
$\mathbf{E}_1\!\parallel \!\mathbf{a}$ polarization above $T_c$ that starts
gradually below  20\,cm$^{-1}$ (Fig.~\ref{Fig_E_ab_4_40}(a))
and extends up to 400\,cm$^{-1}$ (Ref.\cite{Damascelli2000}).
The change of the continuum with temperature is correlated with the 
phase transition at 34\,K as demonstrated in Fig.\,\ref{gap}
where normalized optical conductivity is plotted for three different frequencies.
The continuum absorption at 4.4\,K shows a threshold at 130\,cm$^{-1}$ 
in $\mathbf{E}_1\!\parallel \!\mathbf{a}$ polarization
while at lower frequencies it is similar to the absorption
in  $\mathbf{E}_1\!\parallel \!\mathbf{b}$ polarization (Fig.~\ref{ab_4KGap}).
It is natural to associate the absorption continuum with 
an optical excitation of  two triplets  since the lowest in energy
one triplet excitation is at 65.4\,cm$^{-1}$.

There are two derivative-like phonon absorption lines, one at 91.2 and
the other at 140\,cm$^{-1}$,
in  the $\mathbf{E}_1\!\parallel \!\mathbf{a}$ absorption spectrum
above $T_c$ in Fig.~\ref{Fig_E_ab_4_40}(a).
This line shape, known as the Fano resonance,
comes from the interaction of a discrete level and a continuum of
states\cite{Breit36,Fano61}.
The Fano lineshape\cite{Fano61} does not describe    a normal absorption-like lineshape.
To account for an arbitrary line shape, including the lorentzian line,
 we use an empirical formula\cite{Homes1995} for the dielectric function $\epsilon(\omega)$
where the asymmetry is described by the  phase $\theta$:
\begin{equation}
\epsilon(\omega)=\epsilon_\infty +\frac{ \Omega_p^2 \exp (-\imath\theta )}
        {\omega_p^2-\omega^2-\imath \omega \gamma }.
\label{epsilon}
\end{equation}
$\Omega_p$ is the plasma frequency, $\epsilon_\infty$ is the background 
dielectric constant,  and $\omega_p$ is the
resonance frequency of the phonon, and
$\gamma$ is the full width at half maximum (FWHM) of the lorentzian
line at $\theta = 0$.

We did a $T$ dependence study of the 91\,cm$^{-1}$ line and
the evolution of this  line  with $T$ is shown in Fig.\,\ref{FanoWater}.
The phonon line  at 140\,cm$^{-1}$ is on a steeply rising background.
Since the frequency dependence of the absorption background is not known
it is difficult to subtract the background reliably and
we did not attempt to evaluate the $T$ dependence of the 140\,cm$^{-1}$
line parameters.
The absorption lines were fit with Eq.\,\ref{epsilon}. 
The relation between the real part of the conductivity and dielectric function is
$\sigma_1(\omega) \! = \! \omega \,\mathrm{Im} [ \epsilon(\omega) ]\,/60$, where
$\sigma_1 $ is in units $\Omega^{-1}\mathrm{cm}^{-1}$ and Eq.\,\ref{eq:alpha} 
was used to calculate $\sigma_1$ from the measured absorption spectra.
The fit parameters are plotted in Fig.~\ref{Fanofit}.
The phase $\theta$ (Fig.~\ref{Fanofit}(d))
changes  from $3\pi/2$ at 35\,K to $\pi$ at 33\,K
where the line has the shape of an antiresonance (Fig.~\ref{FanoWater}).
At 31\,K the line has a derivative-like lineshape again,
but with a phase $\pi/2$.
Also the phonon resonance frequency $\omega_p$ stays fairly constant at
higher temperatures and changes abruptly at 35\,K.
The line area $S$ and $\gamma$ have a smooth $T$ dependence around $T_c=35$K.
At 4.4\,K  $\theta =0$ and the normal lineshape is recovered.
The line width $\gamma$ (Fig.~\ref{Fanofit}(a)) goes down with $T$ and
is limited by the 0.4\,cm$^{-1}$  instrumental resolution for this $T$ dependence study.
We know from the higher  resolution measurement  that the line width is 0.2\,cm$^{-1}$  
or less at 4.4\,K.
The line area (Fig.~\ref{Fanofit}(c)) has a different $T$
dependence than other parameters, having a minimum at
approximately 32\,K.
$S$ changes substantially even above $T_c$, by a factor of 2 from 35 to 40\,K.

In this paper we will not present a theory covering the optical conductivity 
of a system where  phonons interact with  a two-particle continuum 
of magnetic excitations.
A proper theory must account for a microscopic mechanism responsible 
for the optical absorption continuum\cite{ChargedMagnon}. 
Nevertheless, some observations can be made based on the empirical fit
of the phonon line shape.
The phase $\theta$ that is related to the Fano parameter $q$,
$q^{-1}\propto \tan (\theta/2)$,
depends on the strength of the interaction between the phonon and the magnetic system.
 The $T$ dependence of $\theta$ shows that the spin-phonon interaction weakens as $T$ 
is lowered below 35\,K. 
The other reason why  the normal lineshape of  the 91\,cm$^{-1}$ phonon at 
low $T$ is recovered could be vanishing of the two particle absorption  continuum
below 130\,cm$^{-1}$.
The direct evidence that the spin-phonon interaction is switched off 
or is very weak at low temperatures
is the lorentzian shape of the 140\,cm$^{-1}$ phonon line.

In conclusion, our data shows that there is an absorption continuum  in the a-axis polarized 
optical absorption that develops a gap at low $T$. 
This gap,  130\,cm$^{-1}$, is equal to  twice  the singlet-triplet excitation energy, 65.4\,cm$^{-1}$.
Therefore the absorption continuum can be assigned to an absorption 
of a photon with a simultaneous creation of two magnetic (spin) excitations.
Two low frequency a-axis polarized optical phonons, at 91 and 140\,cm$^{-1}$
interact with the magnetic system
as evidenced by their derivative-like line shape in the high $T$ phase.

\section{Conclusions}

Using far-infrared spectroscopy we have probed spin excitations and 
phonons in the quarter-filled spin ladder compound $\alpha'$-NaV$_2$O$_5$.
The interaction between the spins and the phonons is observed in the gapped spin state 
below 34\,K and above, in the paramagnetic phase.

The zigzag charge order within the ladders in the gapped state
is in accordance with the polarization dependence 
of several infrared active zone-folded optical phonons.
In the high $T$ phase two a-axis optical phonons interact with the continuum of 
excitations as manifested by the Fano line-shape of phonon lines. 
At low $T$ the continuum absorption is gapped with  a threshold energy 130\,cm$^{-1}$,
twice the singlet-triplet gap, and the interaction between the phonons and the continuum of excitations 
is turned off.

The strength of the singlet to triplet absorption at 65.4\,cm$^{-1}$  
is strongly anisotropic. 
Absorption is strongest when the electric field of the incident light
is polarized along the ladder rungs ($\mathbf{E}_1\parallel \mathbf{a}$).
In this polarization the strength of the singlet to triplet absorption
has a weak magnetic field dependence up to 28\,T.
In $\mathbf{E}_1\parallel \mathbf{c}$ polarization a strong magnetic field dependence
of the singlet-triplet absorption is observed.
This field dependence is due to the dynamic DM interaction created by the 68\,cm$^{-1}$
c-axis optical phonon.
In the case of the dynamic DM absorption mechanism  the
singlet-triplet absorption is electric dipole active 
and the polarization of the transition is determined by 
the polarization of the optical phonon creating the 
dynamic DM interaction by the lattice deformation.
Using the presented theory we calculated  
the strength of the dynamic DM interaction.
The dynamic DM interaction, $qD_Q=0.13$\,cm$^{-1}$,
created by the 68\,cm$^{-1}$ c-axis optical phonon is inter-ladder between the spins  
 in the neighboring planes and points along the b-axis.
We assign the strong a-axis polarized absorption also to a
dynamic DM effect, in this case due to the lattice deformation caused by one of the high 
frequency a-axis optical phonons.   
This dynamic DM interaction is intra-ladder and  is along  the c-axis.
Above 28\,T an increase in the S to T$_+$ a-axis absorption intensity is observed.
The origin of the mechanism responsible for that increase is not clear and 
additional measurements above 33\,T are required.

To summarize, the optical singlet to triplet transition in $\alpha'$-NaV$_2$O$_5$
is dominated by an electric dipole active mechanism.
We have observed the resonant enhancement of 
the singlet to triplet transition close to 
the c-axis 68\,cm$^{-1}$ mode.
We described the enhancement of the electric dipole transition with
the theory of the dynamic DM mechanism
and assigned the 68\,cm$^{-1}$  mode to a c-axis optical phonon.
From the analysis of the phonon infrared spectra we concluded that two types 
of ladder planes exist
with the zigzag charge order along the ladders where the charged stripes 
across the ladders are aligned approximately 
in $[110]$ or in $[\bar110]$ directions.

\section{Acknowledgments}
We thank G.\,Blumberg and O.\,C{\'e}pas for helpful discussions.
Work in Tallinn was supported by the Estonian Science Foundation grants 3443, 4926, and 4927.
A portion of the work was performed at the National 
High Magnetic Field Laboratory, which is supported by NSF Cooperative 
Agreement No. DMR-0084173 and by the State of Florida.
U. Nagel was supported by a NATO expert visit grant PST.EV.978692.

We thank a referee of this paper for pointing out that without the Shekhtman correction
the DM term lowers the three-fold degeneracy of the triplet state\cite{Shekhtman1992,Shekhtman1993}.

\bibliographystyle{apsrev}


\begin{thebibliography}{52}
\bibitem[{Cha()}]{ChargedMagnon}
\bibinfo{note}{It has been shown by Damascelli { \em{et
  al.}}\cite{Damascelli2000} that the direct two-magnon optical absorption is
  possible because of the asymmetrical charge distribution on the ladder
  rungs}.
\end{thebibliography}

\end{document}